\documentclass[10pt,journal,final,a4paper,twoside,twocolumn]{IEEEtran}
\usepackage{float}

\usepackage{dsfont}
\usepackage{amssymb}
\usepackage{amsfonts}
\usepackage{amsmath}
\usepackage[all,poly]{xy}
\usepackage{multirow}
\usepackage{algorithm}
\usepackage{algorithmic}
\usepackage{color}
\usepackage[noadjust]{cite}
\usepackage{subfigure}
\usepackage{tikz,pgf}
\usepackage{array}
\usepackage[colorinlistoftodos]{todonotes}
\usetikzlibrary{fit}
\usetikzlibrary{arrows,automata}
\usepackage{verbatim}
\usetikzlibrary{positioning,shapes.geometric}
\usepackage{url}
\usepackage[T1]{fontenc}

\newcommand{\figwidth}{\columnwidth}

\def\bsd{{\boldsymbol{d}}}

\def\bsm{{\boldsymbol{m}}}

\def\bst{{\boldsymbol{t}}}
\def\bsu{{\boldsymbol{u}}}

\def\bsy{{\boldsymbol{y}}}

\def\bsB{{\boldsymbol{B}}}

\def\bsY{{\boldsymbol{Y}}}

\newcounter{algo}
\renewcommand{\thealgo}{\arabic{algo}}

\newcommand{\beps}{\boldsymbol{\epsilon}}
\newcommand{\bthe}{\boldsymbol{\theta}}

\makeatletter
\newcommand{\thickhline}{%
    \noalign {\ifnum 0=`}\fi \hrule height 1pt
    \futurelet \reserved@a \@xhline
}
\newcolumntype{"}{@{\hskip\tabcolsep\vrule width 1pt\hskip\tabcolsep}}
\makeatother

\title{Fast Task-Based Adaptive Sampling for 3D Single-Photon Multispectral Lidar Data}

\author{Mohamed Amir Alaa Belmekki, \thanks{School of Engineering and Physical Sciences, Heriot-Watt University, Edinburgh, 
EH14 4AS, United Kingdom.}Rachael Tobin, Gerald S. Buller, Stephen McLaughlin, Abderrahim Halimi, \thanks{This work was supported by the UK Royal Academy of Engineering under the Research Fellowship Scheme (RF/201718/17128), and EPSRC Grants 
EP/T00097X/1, EP/S000631/1, EP/S026428/1.}
\vspace{0.2cm} \\ 
  School of Engineering and Physical Sciences, Heriot-Watt University, Edinburgh U.K. \\
   e-mail: $\{$mb219, R.Tobin, G.S.Buller, s.mclaughlin, a.halimi$\}$@hw.ac.uk \vspace{-0.05cm} \\
   }

\begin{document}
\maketitle 
\begin{abstract}  

3D single-photon LiDAR imaging plays an important role in numerous applications. However, long acquisition times and significant data volumes present a challenge to LiDAR imaging. This paper proposes a task-optimized adaptive sampling framework that enables fast acquisition and processing of high-dimensional single-photon LiDAR data. Given a task of interest, the iterative sampling strategy targets the most informative regions of a scene which are defined as those minimizing parameter uncertainties. The task is performed by considering a Bayesian model that is carefully built to allow fast per-pixel computations while delivering parameter estimates with quantified uncertainties.  The framework is demonstrated on multispectral 3D single-photon LiDAR imaging when considering object classification and/or target detection as tasks. It is also analysed for both sequential and parallel scanning modes for different detector array sizes. Results on simulated and real data show the benefit of the proposed optimized sampling strategy when compared to fixed sampling strategies.
\end{abstract} 
 
\section{Introduction} \label{sec:Introduction} 
 
Light detection and ranging (LiDAR) used with time-correlated single-photon detection is receiving significant interest as an emerging approach in numerous applications such as Defence, automotive \cite{WallaceHalimi,rapp2020advances},  environmental sciences \cite{wallace2013design}, long-range depth imaging \cite{tan2020deep,li2020super,mccarthy2009long,li2020single,pawlikowska2017single,li2017multi}, underwater \cite{maccarone2015underwater,halimi2017object} or  through fog \cite{tobin2019three} depth imaging, and multispectral imaging \cite{altmann2016efficient,halimi2019robust}. 
Such a single-photon LiDAR system operates by illuminating the scene using a pulsed laser source, and recording the arrival times of the reflected photons with respect to the time of laser pulse emission. By performing this measurement repeatedly over many laser pulses, it is possible to form a timing histogram from which a high resolution measurement of the photon time-of-flight can be made. By measuring the time-of-flight at each pixel location, it is then possible to make a depth estimate at each part of the scene. This can be done by scanning pixels sequentially, or more efficiently by considering state-of-the-art Single-Photon Avalanche Diode (SPAD) detector arrays which allow the parallel photon acquisition of all pixels \cite{Gyongy_Optica20}.
The resulting histograms of counts contain useful information regarding the presence/absence of an object, and allow a 3D model of the observed target to be built using its estimated depth and reflectivity profiles. This process can be repeated using different laser wavelengths to obtain additional multispectral information on the observed scene. Current limitations to allow extensive use of LiDAR in real world applications include the high acquisition time necessary to collect enough target photons, in addition to the high background level when imaging in ambient light conditions, which affects the quality of the reconstructed 3D scene. Multispectral LiDAR can lead to large data volumes which highlighting the need to reduce measurement points and only target informative regions. Indeed, reducing the target's laser exposition is also important due to eye safety constraints for autonomous navigation, as well as important for energy efficiency systems. Minimising illumination levels is also important when considering medical applications such as single-photon
microscopy in which the observed cells are light sensitive and subject to photostructural damage. 

\begin{figure}
\centering

  \includegraphics[width=9cm,height=9cm]{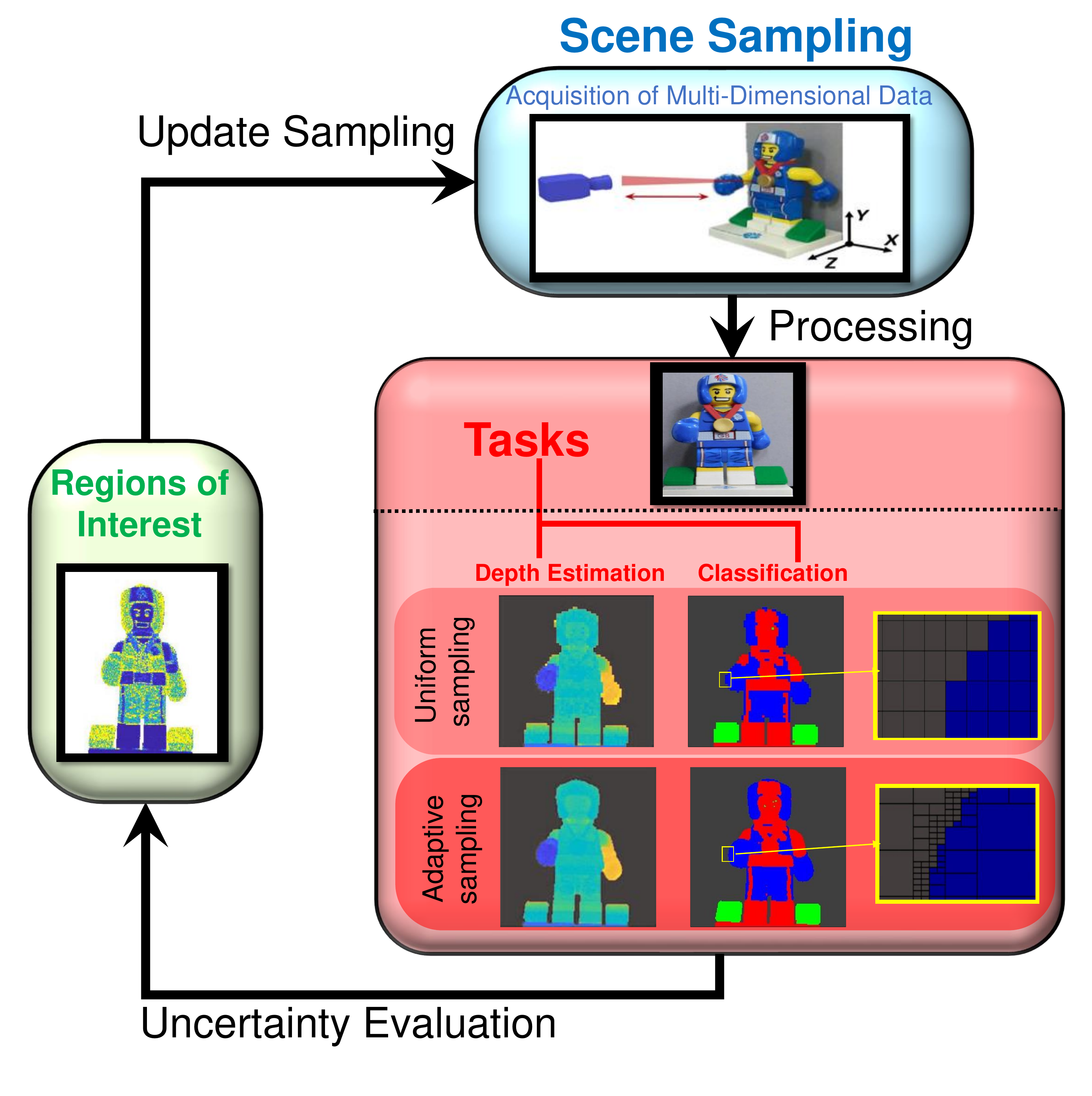}
  \caption{Flow chart representing the proposed adaptive sampling process. }
\label{fig:AS_FChart}
\end{figure}

Many strategies have been proposed to circumvent the above challenges either by improving the acquisition \cite{bergman2020deep,phillips2017adaptive,tobin2017comparative,halimi2019fast,tachella2019bayesian}, or by presenting advanced algorithms to restore damaged or sparse photon data \cite{rapp2017few,halimi2019robust,halimirestoration,rapp2021high} of multidimensional
single-photon LiDAR images. 
In \cite{halimi2019fast}, a dynamic sampling scheme was proposed which showed significant improvement in depth restoration over static sampling schemes. However, the algorithm simplifies the depth reconstruction by assuming that the contribution from ambient light photons is negligible compared to return signal photons.
The methods introduced in \cite{tachella2019bayesian,tobin2017comparative} showed satisfactory results under challenging conditions. Nonetheless, they are still incompatible with real-time requirements and are computationally expensive as they use reversible-jump Markov chain Monte Carlo (RJ-MCMC) as an inference tool.

This paper proposes a new framework for task-optimized adaptive sampling (AS) of the scene to jointly improve both the acquisition and processing of single-photon sensing systems. Based on a selected task, the data acquisition step can be optimized by dynamically sampling the most informative locations of a scene as in   \cite{halimi2019fast,godaliyadda2017framework}. As indicated in Fig. \ref{fig:AS_FChart}, the proposed scene-based sampling strategy is based on three main steps, (i) select the points to scan and their dwell time, (ii) use scanned points to perform a task using a statistical framework with uncertainty quantification, (iii) construct a map of regions of interest (ROI) to define the next  set of scanning points. In this paper, we demonstrate this framework on a LiDAR 3D imaging application and focus on the task of target detection based on object spectral signatures.  
A Bayesian framework is adopted to perform this task since (i) it allows regularization of the ill-posed problem resulting from the scene sub-sampling, (ii) it benefits from marginalisation tools that lead to fast analytical estimates of the parameters of interest, and (iii) it quantifies parameter uncertainties which will be used to define the ROI map. The proposed model accounts for data Poisson statistics and parameter prior information, to build a posterior distribution of the parameters of interest. These parameters include spatial labels to locate pixels with or without a reflective surface, the class of each pixel based on a known spectral library, and  depth estimates for pixels having an object.    
The proposed AS framework and Bayesian algorithm are analysed when considering different sampling scenarios (pixel wise or array scanning), and validated on sparse data with high background levels. The study shows    promising results when compared to conventional sampling strategies such as uniform sampling (US) and random sampling (RS).

To summarize the main contributions of the paper are: 
\begin{itemize}
\item The use of a scene-based sampling approach applied to multiple tasks in the context of single-photon LiDAR imaging. This sampling modality allows fast data acquisition by sampling the most informative regions to perform a task on a particular scene.
\item A new computationally efficient Bayesian algorithm to perform multispectral classification, depth estimation and target detection.
\end{itemize}

The paper is structured as follows.
Section~\ref{sec:Problem_formulation} introduces the multispectral LiDAR observation model. Section~\ref{sec:Adaptive_sampling} presents the task-based adaptive sampling approach. The Bayesian model for the classification task and the associated estimation strategy are presented in Sections~\ref{sec:HBMdeol} and \ref{sec:ESTRA}, respectively. 
Results on simulated and real data are analysed in Sections  \ref{sec:Results_on_simulated_data} and  \ref{sec:Results_on_real_data}.  The Conclusions and future work are finally reported in Section~\ref{sec:Conclusions}.
 
\begin{figure}[h]
\centering
\includegraphics[width=0.7\figwidth,height=3.4cm]{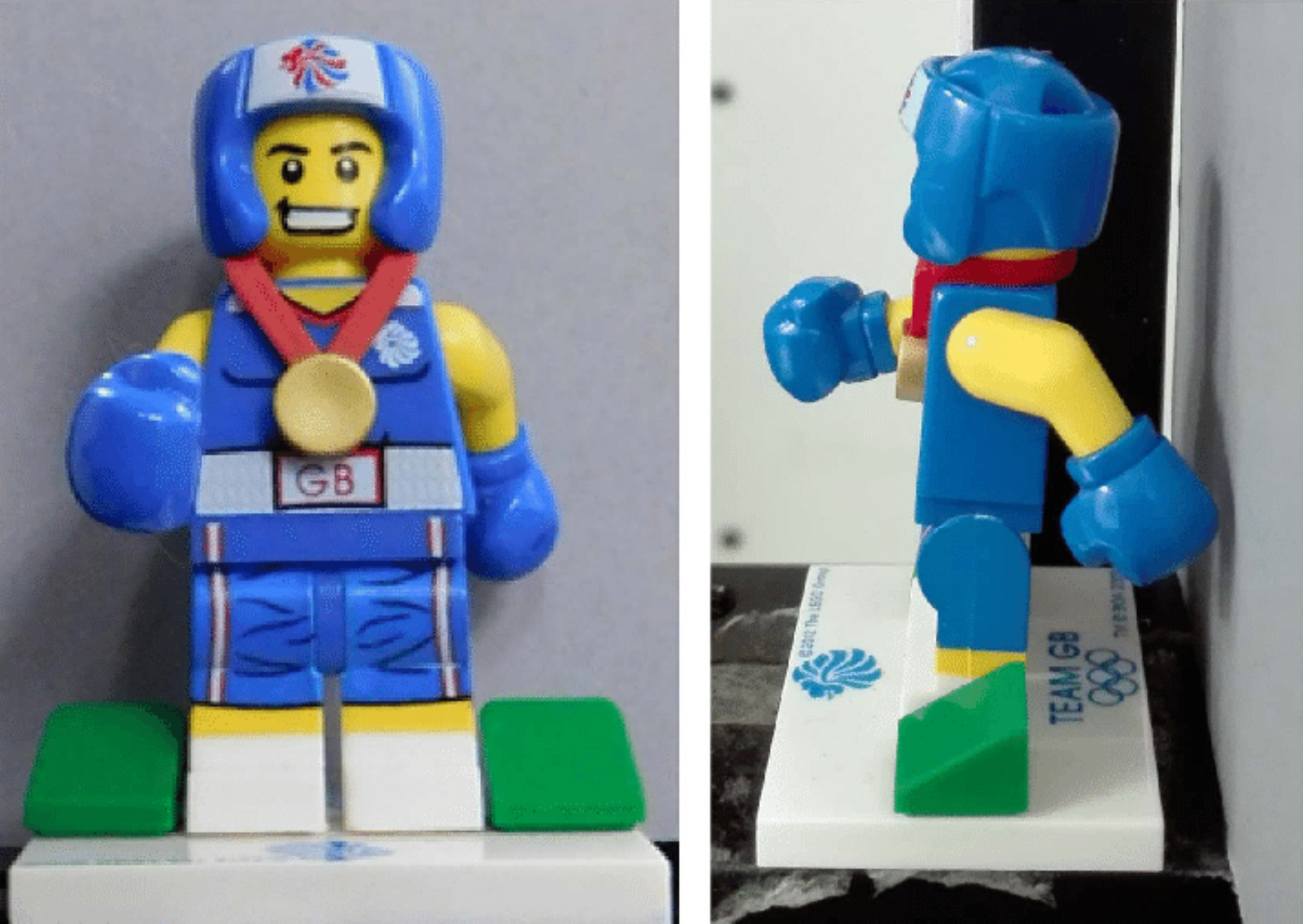}
\vspace{-0.9cm}
\label{fig:IRF}
\end{figure}
\begin{figure}[h]
\centering
\includegraphics[width=0.9\figwidth,height=5.1cm]{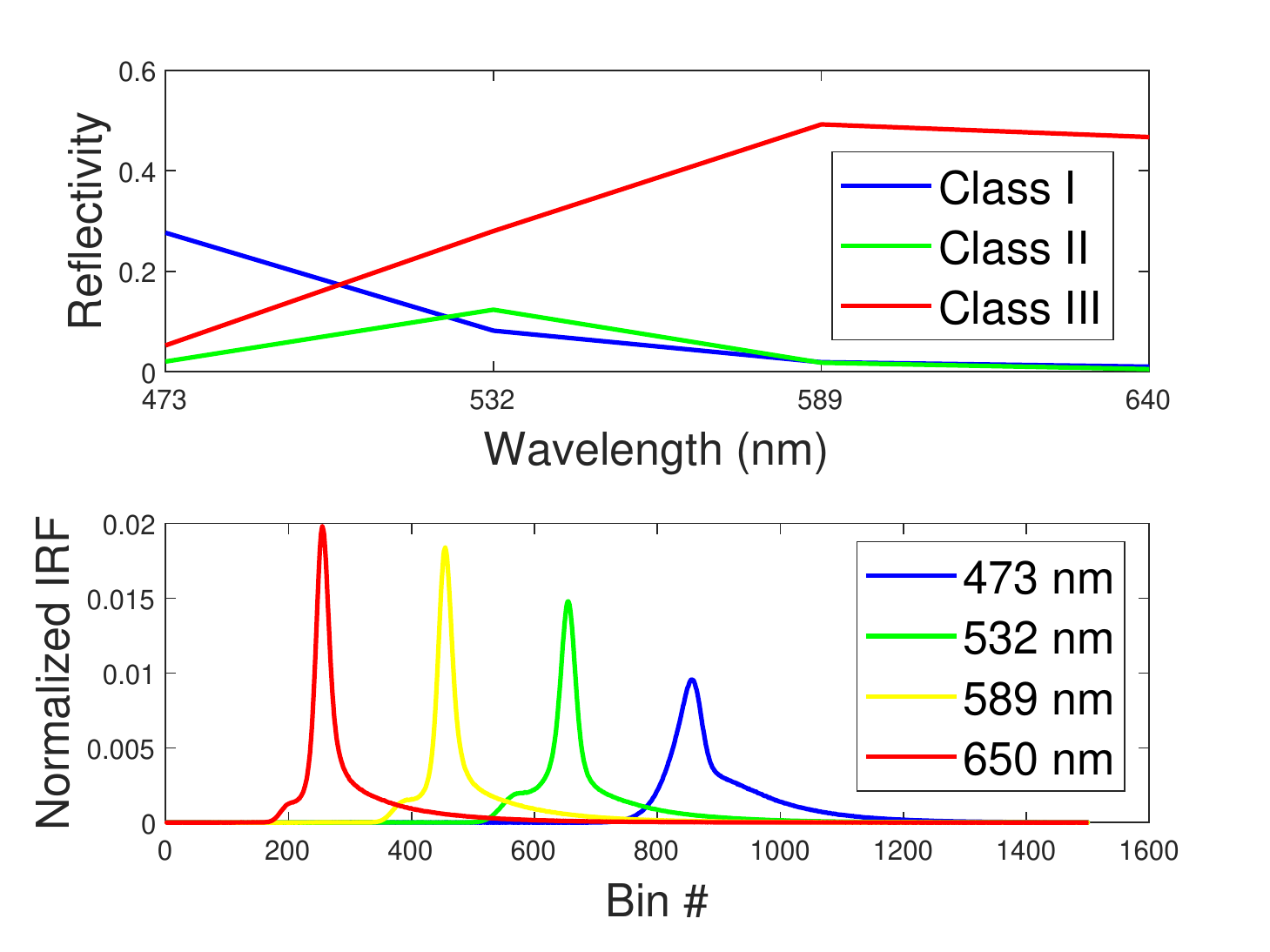}
\vspace{-0.0cm}
\caption{(Top) The considered Lego scene used in section \ref{sec:Results_on_real_data}, (middle) spectral signatures of the spectral classes associated with Lego scene, (bottom) the normalized IRFs associated with the wavelengths 473, 532, 589 and 640 nm \cite{tobin2017comparative} used in Section~\ref{sec:Results_on_real_data}}
\label{fig:IRF}
\end{figure} 
 
\section{Problem formulation} \label{sec:Problem_formulation}

We consider a 3-dimensional cube of histograms $\boldsymbol{Y}$ of LiDAR photon counts of dimension $N \times L \times T$, where $N$, $L$ and $T$ are the number of pixels, spectral wavelengths and  time bins, respectively. Let  $\bsY_{n}=[\bsy_{n,1},\bsy_{n,2},...,\bsy_{n,L}]^T$ be an $L \times T$ matrix where  $\bsy_{n,l}=[y_{n,l,1},y_{n,l,2},...,y_{n,l,T}]^T$.  According to \cite{hernandez2007bayesian, Rapp_TCI2017}, each photon count $y_{n,l,t}$, where ${n \in \{1,...,N\}}$, $l \in \{1,...,L\}$ and $t \in \{1,...,T\}$, is assumed to follow a Poisson distribution given by:
\vspace{-0.0cm}
\begin{equation} \label{eq:4.1}
{y}_{n,l,t} | {r}_{n,l}, d_{n},  {b}_{n,l} \sim \mathcal{P}[{r}_{n,l}\,g_{l}(t-d_{n}) +{b}_{n,l}],
\end{equation}
where $\mathcal{P(.)}$ denotes a Poisson distribution, $r_{n,l} \geq 0$ is the reflectivity observed at the $l$th
wavelength, $d_{n} \in \{1,2,...,T\}$ represents the position of an object surface at a
given range from the sensor, $b_{n,l} \geq 0$ is the constant
background level associated with dark counts and ambient illumination and $g_{l}(.)$ is the system impulse response function (IRF), whose shape can differ between wavelength channels (see Fig.\ref{fig:IRF}), assumed to be known from a calibration step and normalized $\sum_{t=1}^Tg_{l}(t)=1$.
An equivalent model can be considered as in \cite{tachella2019fast} using the signal-to-background ratio (SBR), defined as the ratio of the  useful detected photons $r_{n,l}$ and the total number of background photons in the histogram $b_{n,l}T$, i.e.: $w_{n,l} = \frac{r_{n,l}}{b_{n,l}T}$ with $w_{n,l} \geq 0$. Thus, (\ref{eq:4.1}) can be written in the following form
\begin{equation} \label{eq:4.2}
\mathit{y}_{n,l,t} | {\omega_{n,l}}, d_{n}, b_{n,l} \sim \mathcal{P}\{\mathit{b}_{n,l}\,[w_{n,l} \, T \, g_l(t-d_{n})+1]\}.
\end{equation}
This formulation allows an easy marginalization of the posterior distribution with respect to (w.r.t.)  the background noise parameter as indicated in Section ~\ref{sec:HBMdeol}. 
The joint likelihood, when assuming the observed pixels, wavelengths and bins mutually independent, is then given by
\begin{equation} \label{eq:4.3}
{\mathit p(\bsY | \boldsymbol{\Omega},  \bsd, \bsB) = \prod_{l=1}^{L} \prod_{t=1}^{T} p({y_{n,l,t}} | {\omega_{n,l}}, d_{n}, b_{n,l})
}
\end{equation}
where $\bsd= \left(d_1,\cdots, d_N \right)$ and  $\boldsymbol{\Omega}$, $\bsB$ are two  matrices gathering $\omega_{n,l}, \forall n,l$, and $b_{n,l}, \forall n,l$, respectively.
Considering our observation model, it is clear that the data generated becomes computationally expensive as we increase the time bin and the spatial and/or the spectral resolution. For instance, a data set with $100 \times 100$ pixels, $1000$ time bins, and $32$ wavelengths can yield an excessively large number of data samples ( >$10^8$) which will lead to a significant computational load and prohibitive memory requirements.
To circumvent the data limitation aspect of our model, the following section will propose a scene-based sampling strategy for smart and informed data acquisition.


\vspace{-0.0cm}
\section{Task-optimized adaptive sampling} \label{sec:Adaptive_sampling}
\vspace{-0.0cm}


The proposed strategy is summarized in Fig. \ref{fig:AS_FChart}. Assuming the presence of a high-resolution (HR) sampling grid of $N$ pixels, the approach aims to iteratively sample $N_s$ pixels to progressively improve the performance of a pre-defined task (e.g., depth estimation, classification, etc). The location of these $N_s$ pixels is subject to different constraints based on the scanning scenario considered. Indeed, the $N_s$ pixels could be non-uniformly located in the scene when considering pixel-wise scanning of the scene, or have a spatial structure when considering an array scanning system. Both scenarios will be investigated in the following section. The collected samples will then be used to perform specific tasks such as target detection, classification, depth estimation, etc. The uncertainty of the estimated parameters are then used to establish a map of regions of interest that will serve as a basis for defining the positions and the acquisition times of the new sampled pixels. This iterative procedure will continue until convergence, as indicated in Algo. \ref{alg:S_B_A_S} where $i$ represents the current iteration. The next subsections describe in detail each of these steps.

\begin{algorithm}

\def\NoNumber#1{{\def\alglinenumber##1{}\State #1}\addtocounter{ALG@line}{-1}}

\caption{Task-optimized adaptive sampling} \label{alg:S_B_A_S}
\begin{algorithmic}[1]
       \STATE \underline{Initialization}
       \STATE Initialize: $N_s$, pixel locations  $ \nu_{N_s}^{(1)}$, pixels acquisition times are $t_{\nu_{N_s}}^{(1)}$, $N$,  conv=0  
       \WHILE{conv$=0$}
       
                \STATE Scan $N_s$ pixels with locations $\nu_{N_s}^{(i)}$  and acquisition times $t_{\nu_{N_s}}^{(i)}$ \\
                \STATE \underline{Processing step:}     \\	
	            
	            Fast and robust task performance (object detection, classification, depth estimation, etc)
				
				\STATE \underline{Regions of interest:}  
				
				Computation of a probability map of ROI \\
				
				\STATE \underline{Update sampling step:} 
				
				Generation of new positions  $ \nu_{N_s}^{(i)}$ (using a MH algorithm) and acquisition times $\bst_{\nu_{N_s}}^{(i)}$ 
			 
				\STATE \underline{Convergence:}  
				
    		    conv$= 1$, if the stopping criteria are satisfied.  
       \ENDWHILE 
\end{algorithmic}
\end{algorithm}

\subsection{Fast and robust task performance} \label{subsec:Parameter_estimation}

Sensing aims to collect information regarding some phenomena and to perform a pre-defined task, such as depth estimation in LiDAR, cell detection in microscopy, target tracking in defence, etc. This step of the algorithm performs this pre-defined task by solving an inverse problem. In this work, we adopt a Bayesian approach that leads to the estimation of $M$ parameters of interest for the $n$th scanned position, denoted $\bthe_n = \left(\theta_{n,1}, \cdots, \theta_{n,M} \right)$, together with a measure of their uncertainties $\beps_n = \left(\epsilon_{n,1} , \cdots, \epsilon_{n,M}  \right)$, where both $\bthe_n $ and $\beps_n$ are extracted from the estimated parameter posterior distributions. These estimates are only available on scanned points, which represent a subset of the high-resolution grid composed of $N$ pixels. To obtain a high-resolution ROI map, the obtained sub-sampled estimates should be spatially extended to cover more space in the HR grid. 
This leads to an inpainting problem, in which the parameters and uncertainties of an $n$th unavailable pixel (i.e., unscanned pixels or scanned pixels without photon detection) can be inferred from scanned neighbours pixels, as follows $\bthe_n  = h_p\left(\bthe_{\psi_n}\right)$, and  $\beps_n   = h_u\left(\beps_{\psi_n}\right)$
where $\psi_n$ represents the   $3^{\textrm{Wind}} \times 3^{\textrm{Wind}} $ window of neighbours of the $n$th location, and $h_p$, $h_u$ represent non-linear inpainting operators on the parameters and uncertainties, respectively.
In this paper, we are interested in fast operators allowing efficient AS iterations, and thus adopt  the median filter for both $h_p$, $h_u$ in what follows. Other advanced optimization or learning based algorithms could be considered, which is beyond the scope of this paper. 
The resulting HR estimates and uncertainties will be used to build the region of interest (ROI) map, as described in the following section.


\subsection{Regions of interest} \label{subsec:Regions_of_interest}

%
In this subsection the algorithm will generate a probability map $\bsm \in \mathds{R}^N$ (with $m_n\geq0$ and $\sum_{n} m_n=1$) of regions of interest, where regions with higher values should be scanned more frequently. 
This ROI is closely related to the targeted task and the nature of the parameters of interest. 
Based on an estimation task, the method in \cite{halimi2019fast} aimed to improve depth estimates and defined the ROI based on depth gradients, while the method in \cite{godaliyadda2017framework} introduced additional features based on the distance between scanned points. In a target detection scenario, where a target is defined as the presence of a reflective surface in the single-wavelength case or  an object with a specific spectral signature in the multispectral case, the AS procedure should focus on scanning regions having a target and spending less resources on other regions of the scene. 
This paper assumes the presence of uncertainty measures $\beps$ in addition to parameter estimates $\bthe$, and propose to build the ROI as follows 
$\bsm = h_r \left(\bthe, \beps\right)$, where $h_r$ is a chosen nonlinear function. The latter formulation allows the combination of multiple tasks, and to take advantage of the available uncertainty measures.  For example, multiple tasks could happen if we are interested in reconstructing the depth profile of an object with a specific spectral signature. In this case, the parameters  $\bthe$ could gather a depth estimate and a spatial label that classifies pixels based on their spectral signatures,  and the ROI map should highlight pixels belonging to that object and give a particular interest to pixels with high depth uncertainty.
Finally, it should be noted that pixels with no estimates after the inpainting process have the highest uncertainty and should receive high  probability of sampling, in contrast to pixels that reached maximum acquisition time which should be excluded from the next sampling iterations.

\subsection{Generation of new locations and acquisition times} \label{subsec:sample_generation}
\begin{figure}[h]            
\includegraphics[width=1\figwidth]{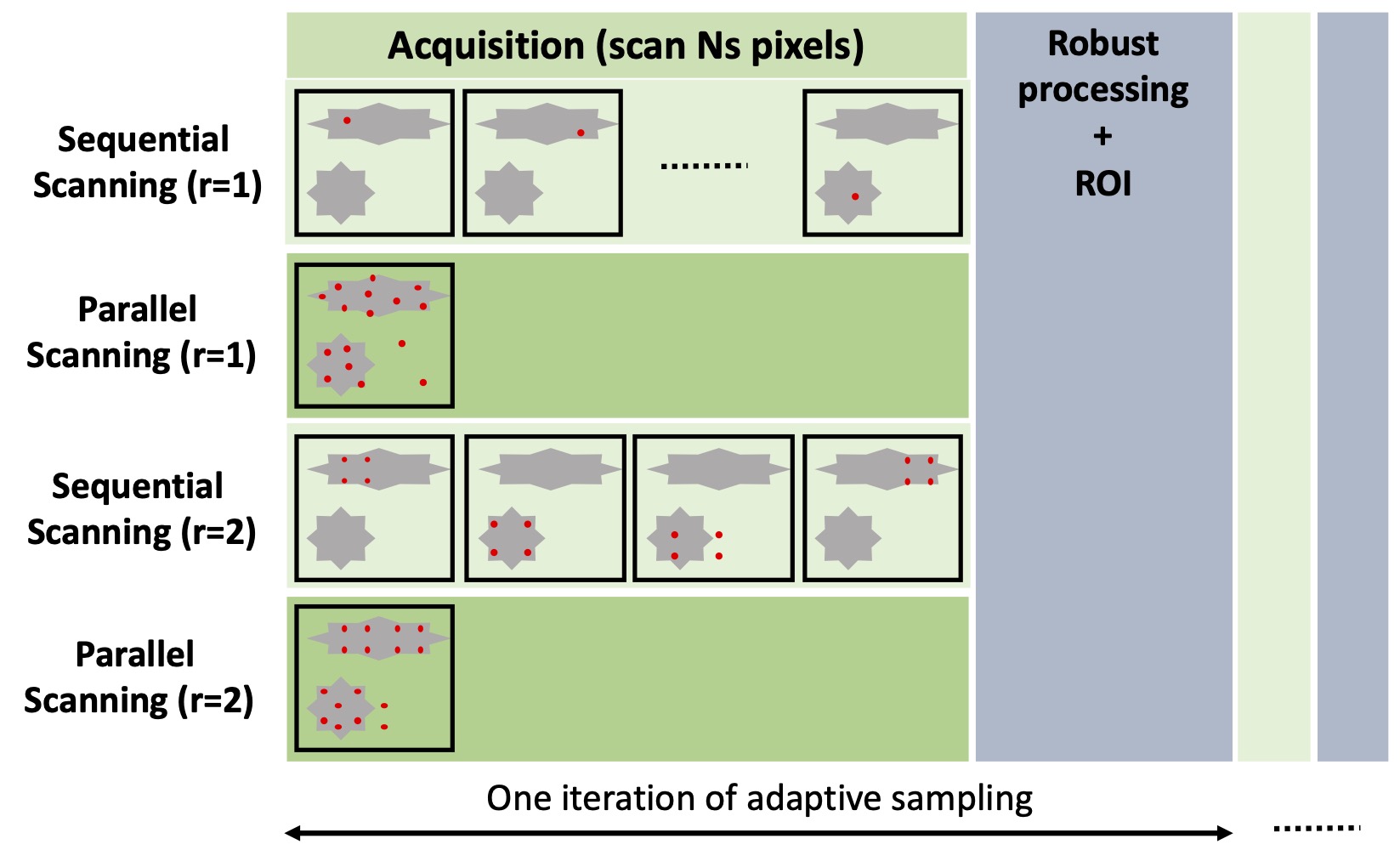}
\caption{Illustration of different acquisition scenarios where gray objects represent regions of interest.  $N_s$ pixels (red dots) are scanned in each iteration of adaptive sampling ($N_s=16$ in this example) using sequential or parallel scanning, with different array sizes ($r=1$ and $r=2$). Some pixels are outside the ROI to explore all parts of the scene.}
\label{fig:Scan_scenarios}
\end{figure}

The proposed strategy is based on sampling  $N_s$ pixels at each adaptive sampling iteration. The locations of scanned points are subject to physical constraints due to the sampling scenario, i.e.,  pixel-wise scanning or array based sampling as shown in Fig. \ref{fig:Scan_scenarios}. 
Single-photon pixel-wise scanning has been widely used as it allows eye-safe imaging even at km-range distances \cite{altmann2016lidar,mccarthy2013kilometer,tobin2017comparative}. With recent technological advances, array scanning is made possible by moving a detector array spatially to scan a large field of view \cite{martin2021high,wade2021sub}. In this paper, we study both scanning cases, and investigate the scenarios of sequential scanning (i.e., when scanning one pixel/array at a time) and parallel scanning when scanning all pixels in one shot (e.g., using a binary mask to select pixels or by some pixels of an array of detectors), as illustrated in  Fig. \ref{fig:Scan_scenarios}. 
Given the ROI probability map, we can select the next locations to scan ($N_s/r^2$ locations when scanning using $r \times r$ array, i.e., $N_s$ locations in pixel-wise scanning) by selecting those with the highest probabilities. As stated in \cite{halimi2019fast}, this strategy prevents the exploration of the full scene which might lead to missing small or dynamic objects. Consequently, given the ROI probability map, the new locations will be randomly sampled according to  the distribution $\bsm$ using the Metropolis-Hastings algorithm \cite{casella1999monte} with a uniform proposal distribution on the HR grid. It is worth noting that array scanning requires   selecting $N_s/r^2$ array positions together with the associated array sizes, i.e., distance between points that mimic an array zoom-in or zoom-out. These array sizes are fixed using multi-scale information of the ROI map $\bsm$. Finally, it can be seen from Fig. \ref{fig:Scan_scenarios} that the dwell time depends on the scenario and the acquisition time per shot. For example, consider a fixed acquisition time per step $t_0$, the sequential array scanning will require $N_s t_0/r^2$, while parallel scanning could be done in one shot requiring $t_0$ acquisition time.   
Each new location (to be sampled from $N_s$ or $N_s/r^2$) will be assigned an acquisition time ranging from $t_0$ to $ct_0$, where $t_0$ is a user defined acquisition time step and $c$ is the importance level of a shot. During the iterative process, the defined acquisition time step is updated to ensure that the proportion of scanned pixels with photon detections is between $[0.7, 0.9]$. This will avoid working with too small time steps leading to low photon detections, or wasting resources by using unnecessary long time steps. 

\subsection{Stopping criteria} \label{subsec:classif_strat}

Many stopping criteria can be considered for Algo. \ref{alg:S_B_A_S}. The first criterion compares the two last depth parameter estimates and stops the algorithm if their root mean square error is smaller than a given user-defined threshold, that is:  $\textrm{RMSE} (d^{(t+1)},d^{(t)})  \leq \xi$, where $\textrm{RMSE} (d^{(t+1)},d^{(t)})$ denotes the root mean square error between the depth estimates at the iteration $t+1$ and $t$.
Other criteria can be considered such as reaching a pre-defined maximum acquisition time-per-pixel, a maximum number of scanned points and/or a maximum number of iterations.

\section{Hierarchical Bayesian model for classification strategy} \label{sec:HBMdeol} 
The proposed adaptive sampling strategy is optimized for a predefined task. In this paper, we are interested in a target detection task based on a known object spectral signature. More precisely, we propose a spatial classification algorithm that labels pixels based on their spectral signatures, which then allows concentrating scanning samples on pixels of interest. Due to data sub-sampling during acquisition, the resulting histograms are often sparse and some pixels might be empty, leading to an under-determined problem to perform the task. The latter is solved using a Bayesian approach that performs the classification task on multispectral LiDAR data. Adopting a Bayesian framework, the unknown parameters  will be assigned prior distributions that will allow inclusion of additional information and regularization of the ill-posed problem. In addition to parameter estimates, this approach will also provide uncertainty measures regarding the estimated parameters, as required by the AS strategy. 
 
 
 
 
\subsection{Prior distributions} \label{subsec:Prior distributions}

A LiDAR histogram can either result from background counts (in the absence of a target photons due to $r_{n}=\omega_{n} = 0$) or a mixture of target and background counts (when $r_{n} \geq 0$ or $\omega_{n} \geq 0$). Assuming the presence of $K$ spectral signatures, the classification problem aims to associate a pixel with a target to one of the $K$ spectral classes. The reflectivity prior accounts for this effect by considering a mixture of $K+1$ distributions as follows 
\begin{eqnarray} \label{eq:4.4}
\mathit P(r_{n,l}|u_n,\alpha_{k,l}^{r},\beta_{k,l}^{r},K) 
= \delta {(u_n)} \delta {(r_{n,l})}  \nonumber \\ 
+ \sum_{k=1}^{K} \delta (u_n-k) \mathcal{G}(r_{n,l};\alpha_{k,l}^{r},\beta_{k,l}^{r}) 
\end{eqnarray}
where $u_n \in \{0,1,...,K\}$ is a latent variable that indicates the absence of target if $u_n=0$, otherwise, it indicates the label of the class,  $\delta(.)$ is the Dirac delta distribution centred in 0,  $\mathcal{G}(r_{n,l};\alpha_{k,l}^{r},\beta_{k,l}^{r})$ represents a gamma density whose shape and scale hyperparameters $\left(\alpha_{k,l}^{r}, \beta_{k,l}^{r}\right)$  are fixed based on the $K$ known spectral signatures. 
This prior is inspired from the spike-and-slab prior used in  \cite{tachella2019fast}. It accounts for $K+1$ cases, the first represents the absence of a target in the $n$th pixel and is obtained for $u_n =r_{n,l} = 0 $, hence the use of a Dirac distribution (the spike part).  The slab part accounts for the presence of one of the $K$ signatures by using a gamma distribution.
Thanks to the use of many wavelengths, this prior  extends the object detection problem in \cite{tachella2019fast} to a class detection problem using the spectral signature of each class. Considering the non-negativity of $b_{n,l}$, $\forall n,l$  and its continuous nature, the background level will be modelled with a gamma distribution as in \cite{altmann2016robust}:

\begin{equation} \label{eq:4.5}
{\mathit P(b_{n,l}|\alpha_{l}^{b},\beta_{l}^{b}) = \mathcal{G}(b_l,\alpha_{l}^{b},\beta_{l}^b}) 
\end{equation}
where $\alpha_{l}^{b}$ and $\beta_{l}^{b}$ are background hyper-parameters. 
Considering that only dozens of distinctive wavelengths will be used, we will consider the channels to be uncorrelated to keep the estimation strategy tractable. Since we are interested in using the model described in (\ref{eq:4.2}) instead of (\ref{eq:4.1}), assuming that the reflectivity and the background noise are independent and by applying a random variable change, the resulting joint prior distribution will yield:

{
\scriptsize
\begin{eqnarray} \label{eq:4.6}
p(\boldsymbol{\omega_{n}},\boldsymbol{b_{n}} | u_n, \boldsymbol{\phi},K) &= &
 \prod_{l=1}^{L} p(\omega_{n,l},b_{n,l} | u_n, \phi_l)
\nonumber  \\
&= &\mathit \prod_{l=1}^{L} \left[
\delta (u_n)\delta (\omega_{n,l}) \mathcal{G}(b_{n,l},\alpha_{l}^{b},\beta_{l}^b) \right.  \\
& +& \left. \sum_{k=1}^{K}\delta (u_n-k)C_{k,l}(\omega_{n,l}) \mathcal{G}(b_{n,l},\alpha_{l,k}^{\dagger},\beta_{l,k}^{\dagger}(\omega_{n,l}) \right]  \nonumber
\end{eqnarray}
}
with
\begin{equation}
\mathit C_{k,l}(\omega_{n,l}) = \frac{(\beta_l^{b})^{\alpha_l^{b}} (\beta_{k,l}^{r})^{\alpha_{k,l}^{r}} T^{\alpha_{k,l}^{r}}}{B(\alpha_{k,l}^{r},\alpha_{l}^{b})} \frac{\omega_{n,l}^{\alpha_{k,l}^{r}-1}}{\beta_{l,k}^{\dagger}(\omega_{n,l})^{\alpha_{l,k}^{\dagger}}}
\nonumber
 \end{equation}
 
\begin{equation}
\mathit \alpha_{l,k}^{\dagger}=\alpha_{l}^{b}+\alpha_{k,l}^{r}
\nonumber
\end{equation}
\begin{equation}
\mathit \beta_{l,k}^{\dagger}(\omega_{n,l})=\beta_{l}^b +\beta_{k,l}^r T \omega_{n,l}
\nonumber
\end{equation}
where $\mathit B(.)$ is the beta function and $\boldsymbol{\phi} = (\phi_1, \phi_2,..., \phi_l)$ with $\phi_l=(\alpha_{k,l}^{r},\beta_{k,l}^{r},\alpha_{l}^{b},\beta_{l}^{b})$, $k\in \{0,...,K\}$.

As we suppose that we have no prior knowledge about a pixel's class, the parameter $u_n$ is assumed to be drawn from a uniform distribution, i.e.: $p(u_n=k)=\frac{1}{K+1}$, where $k\in \{0,...,K\}$. However, This non-informative class prior can be changed in presence of additional information regarding the classes. The depth parameter $d_n$ is assigned a non-informative uniform prior as follows:
 \vspace{-0.0cm}

\begin{equation} \label{eq:4.7}
\mathit p({d}_{n} = t ) =  \frac{1}{T}
 \quad , \forall{t} \in \{1,...,T\}.  
\end{equation}
Nonetheless, this can be modified in case of additional information regarding the target position.

 
 \vspace{-0.0cm}
\subsection{Joint Posterior distribution} \label{subsec:Joint Posterior distribution}
\vspace{-0.0cm}

From the joint likelihood in (\ref{eq:4.3}) and the prior distributions specified in Section \ref{subsec:Prior distributions}, we can obtain the joint posterior distribution for  $\boldsymbol{\omega_{n}}$,$\boldsymbol{b_{n}}$, $d_n$ and $u_n$ given the 3D histograms $\boldsymbol{Y_n}$ and the hyperparameters $\boldsymbol{\phi}$ and $K$. Using Bayes rule and assuming that $d_n$ and $u_n$ are independent from $\boldsymbol{\omega_{n}}$ and $\boldsymbol{b_{n}}$, the joint posterior distribution of the proposed Bayesian model can be formulated in the following form:
\begin{equation} \label{eq:4.8}
{\mathit p({\boldsymbol{\Theta_{n}}}|\boldsymbol{Y_n}, \boldsymbol{\phi}, K) \propto p(\boldsymbol{Y_n}|{\boldsymbol{\Theta_{n}}}) p(\boldsymbol{\Theta_{n}}|\boldsymbol{\phi},K)
}
\end{equation}
where
\begin{equation}
\boldsymbol{\Theta_{n}}=(\boldsymbol{\omega_{n}},\boldsymbol{b_{n}},d_n,u_n)
\nonumber
\end{equation}
\begin{align} \label{eq:Postereq}
p(\boldsymbol{\Theta_{n}}|\boldsymbol{\phi},K) &= p(\boldsymbol{\omega_{n}},\boldsymbol{b_{n}},d_n,u_n|\boldsymbol{\phi},K) \,   \nonumber  \\
&= p(\boldsymbol{\omega_n},\boldsymbol{b_n}|\boldsymbol{\phi},u_n,K)p(d_n)p(u_n). 
\end{align}

\vspace{-0.0cm}
\section{ESTIMATION STRATEGY} 
\label{sec:ESTRA} 
\vspace{0.0cm}
The posterior distribution in \eqref{eq:4.8}  reflects our knowledge of the unknown parameters to be estimated given the photon data and the available prior knowledge. In this paper, we are interested on estimating the depth and label parameters, i.e., $M=2$ and  $\bthe  = (\bsd, \bsu)$. The Bayesian estimator to be considered, both for the depth and class parameter, is the maximum a posteriori (MAP) estimator as in \cite{altmann2017robust,altmann2016efficient}. From   (\ref{eq:4.8}), we marginalize the background noise and signal-to-background parameters to get the joint depth and class marginal probability as follows:

\vspace{-0.0cm}
\begin{equation} \label{eq:4.10}
\mathit  p(u_n,d_n|\boldsymbol{Y_n)}=  \int \int p(\boldsymbol{\omega_n},\boldsymbol{b_n},d_n,u_n|\boldsymbol{Y_n}) d\boldsymbol{b_n} \, d\boldsymbol{\omega_n}. 
\end{equation}

\vspace{-0.0cm}
\subsection{Class estimation} \label{subsec:class_est}
\vspace{0.00cm}
The following decision rule is adopted to determine the pixel label 
\begin{equation} \label{eq:4.9_1}
{\mathit H_{n}=\max_{k=1:K} \, p(u_{n}=k\,| \,\boldsymbol{Y_n})
}
\end{equation}
with
\begin{equation} \label{eq:4.101}
\mathit  p(u_n|\boldsymbol{Y_n)}= \sum_{d_n=1}^{T} \int \int p(\boldsymbol{\omega_n},\boldsymbol{b_n},d_n,u_n|\boldsymbol{Y_n}) d\boldsymbol{b_n} \, d\boldsymbol{\omega_n} 
\end{equation}
where $H_{n}$ represents the class of the $n$th pixel. Note that for $K=1$ and $L=1$, we end up with a target detection decision rule as in \cite{tachella2019fast}. We demonstrate that the marginal probability $p(u_n|y_n)$ is :
\vspace{-0.0cm}
\begin{eqnarray} \label{eq:4.11}
\mathit  p(u_n=0|\boldsymbol{Y_n}) =  \prod_{l=1}^{L} p(u_n=0|\boldsymbol{y_{n,l}})  \nonumber  \\
      = \prod_{l=1}^{L} \frac{p(u_n=0)\Gamma (\bar{y_{n,l}}+\alpha_l^{b})}{(T+\beta_l^{b})^{(\bar{y_{n,l}+\alpha_l^{b})}\, \gamma_l}}
\end{eqnarray}

\begin{eqnarray} \label{eq:4.12}
\mathit  p(u_n=k|\boldsymbol{Y_n}) = \sum_{d_n=1}^{T} \int_{0}^{\infty} \prod_{l=1}^{L}  \left[p(u_n=k) p(d_n)  D_{n,l,k}  \gamma_l^{-1} \right.    \nonumber \\ 
 \left.    \mathit{F_{n,l,k}}(\omega_{n,l},d_n)  d\omega_{n,l} \right]
\end{eqnarray}
with 
\vspace{-0.0cm}
  \begin{equation}
\mathit \gamma_l =\frac{\Gamma (\alpha_l^{b}) }{{(\beta_{l}^{b})}^{\alpha_{l}^{b}}} \prod_{t=1}^{T} y_{n,l,t} !
\nonumber
  \end{equation}
\vspace{-0.0cm}
 \begin{equation}
\mathit{D_{n,l,k}} = 
\frac{\Gamma (\bar{y}_{n,l}+\alpha_l^{b}+\alpha_{k,l}^{r}) (T\beta_{k,l}^{r})^{\alpha_{k,l}^{r}}\,}{\Gamma (\alpha_{k,l}^{r})}
\nonumber
 \end{equation}
 \vspace{-0.0cm}
 {\small{
 \begin{equation} \label{eq:4.13}
\mathit{F_{n,l,k}}(\omega_{n,l},d_n) =
\frac{\exp \{\sum_{t=1}^{T} y_{n,l,t}\ln [\omega_{n,l} \, T g_{l}(t-d_n)+1]\}}{\omega_{n,l}^{1-\alpha_{k,l}^{r}}\{\beta_{l}^{b}+ [T \, (1+\omega_{n,l}(1+\beta_{k,l}^{r}))]\}^{\alpha_{l,k}^{\dagger}+\bar{y}_{n,l}}}  
\end{equation}}}
where $\bar{y}_{n,l}=\sum_{t=1}^{T} {y}_{n,l,t}$.
In the event of no target, we can see that the integral is available in its analytical form thanks to the conjugacy between the model (\ref{eq:4.2}) and the priors (\ref{eq:4.6}). The marginal
distribution in (\ref{eq:4.12}) is, however, intractable in presence of a target. One way to simplify it is to consider that the depth captured is different across all the spectral wavelengths. This simplification improves the tractability of the marginal class probability and will transform (\ref{eq:4.12}) into (\ref{eq:4.14}) as follows:

\begin{eqnarray} \label{eq:4.14}
 \mathit  p(u_n=k|\boldsymbol{Y_n}) = \prod_{l=1}^{L} \sum_{d_{n,l}=1}^{T} \left[p(u_n=k) p(d_{n,l})  D_{n,l,k}  \gamma_l^{-1}  \right. \nonumber \\
\left. \int_{0}^{\infty} \mathit{F_{n,l,k}}(\omega_{n,l},d_{n,l})  d\omega_{n,l} \right].
\end{eqnarray}

The resulting integral with respect to $\omega_{n,l}$ in \eqref{eq:4.14} can be numerically approximated with a quadrature method. The matched filter in \eqref{eq:4.13} can be computed with $ \mathit {\mathcal{O}(T \textrm{log} T )}$ using the fast Fourier transform (FFT) leading to an overall complexity of the integral per-pixel in (\ref{eq:4.14}) given by $ \mathit {\mathcal{O}(KLJT \textrm{log} T )}$, where $K$ is the number of classes considered, $L$ is the number of wavelengths, $J$ is the computational cost of the evaluated integrand and $T$ is the number of the temporal bins.

\subsection{Depth estimation} \label{subsec:depth_est}
\vspace{-0cm}
The depth estimate can be obtain as follows
\begin{equation} \label{eq:4.9_2}
{\mathit {\hat{d}_{n}}=\max_{d=1:T} \, p(d_{n}\,| \,\boldsymbol{Y_{n}})
}
\end{equation} 
where
\vspace{-0.0cm}
\begin{eqnarray}
p(d_{n}|\boldsymbol{Y_{n}}) =  \sum_{k=1}^{K} \prod_{l=1}^{L}  \left[p(u_n=k) p(d_{n})  D_{n,l,k}  \gamma_l^{-1}  \right. \nonumber \\
\left. \int_{0}^{\infty} \mathit{F_{n,l,k}}(\omega_{n,l},d_{n})  d\omega_{n,l} \right]. \label{eq:depth_l_marg}
\end{eqnarray}




Although $\omega$ can be integrated out from   \eqref{eq:depth_l_marg}, this might lead to a high computational cost. In this paper, we choose to estimate the depth given the easily computed marginal map estimate $\omega^{\textrm{map}}_{n,l}$ using the simplified model introduced in Section \ref{subsec:class_est}, leading to 
\begin{equation} \label{eq:4.9_3}
{\mathit {\hat{d}_{n}}=\max_{d=1:T} \, p(d_{n}\,| \,\boldsymbol{Y_{n}}, \boldsymbol{\omega}^{\textrm{map}}_{n} ).
}
\end{equation} 
The proposed approach allows for the evaluation of the full marginal depth posterior. In addition to the depth point estimate, this distribution will allow uncertainty quantification (i.e., to quantify our confidence regarding the estimates). In this paper, we evaluate the depth uncertainty by considering the depth negative log-cumulative marginal posterior around the MAP estimate, i.e.,
$\textrm{NCD} = -\textrm{log} \left[ \sum_{ {\hat{d}_{n}} - \epsilon}^{\hat{d}_{n} + \epsilon} {p(d_{n}\,| \,\boldsymbol{Y_{n}}, \boldsymbol{\omega}^{\textrm{map}}_{n} )} \right]$, where $\epsilon$ is a user fixed constant. Note that a small NCD  indicates a high confidence about the estimate, while a large one would be an indication of low confidence. 


\vspace{-0.0cm}
\section{Results on simulated data} \label{sec:Results_on_simulated_data} 






This section evaluates the performance of the proposed adaptive sampling framework and classification algorithm  on simulated data. We first evaluate the performance of the classification algorithm using simulated multispectral data with $L=4$ wavelengths when varying the signal-to-background ratio (SBR) and the average signal photons per pixel.  Then, we compare different adaptive sampling scenarios based on sequential and parallel scanning modes for different array sizes. All simulations were performed using Matlab 2020a on an Intel Core i7-8700@3.2GHz, 16 GB RAM and results are averaged based on three Monte Carlo realisations.

\subsection{Evaluation criteria}
\label{subsec:Evaluation_criteria}


To evaluate the performance of the proposed algorithm, we use the root mean square error (RMSE) and the accuracy to evaluate the depth estimation and class estimation, respectively. The RMSE is defined as $\textrm{RMSE} =\sqrt{ \frac{1}{N}  ||\bsd^{\textrm{ref}}- \widehat{\bsd}||^2}$, where $\bsd^{\textrm{ref}}$ is obtained from sampling the whole scene with the maximum acquisition time and under a negligible background illumination (see top-middle and bottom-middle of Fig. \ref{fig:MF_GT}). The accuracy in percentage is defined as $\textrm{ACC} =\frac{\textrm{TP}+\textrm{TN}}{\textrm{TP}+\textrm{TN}+\textrm{FP}+\textrm{FN}}$, where $\textrm{TP}$,  $\textrm{TN}$, $\textrm{FP}$ and $\textrm{FN}$ represent: true positive, true negative, false positive and false negative probabilities, respectively. 

\subsection{Datasets} \label{subsec:Datasets}


Two real single-photon data sets will be used to assess the performance of the proposed algorithm. The first one is the  life-sized mannequin head shown in Fig. \ref{fig:MF_GT} (Top-left image). This data was acquired at a standoff distance of 40 m using a time-of-flight scanning sensor, based on a time-correlated single-photon counting (TCSPC) (the reader is encouraged to read \cite{altmann2016lidar,mccarthy2013kilometer} for more information regarding the transceiver system and data acquisition hardware used for this work). The spatial, spectral and temporal dimensions for the mannequin head are $N=142\times142$ pixels, $L=1$ wavelength and $T=191$ bins (bin width of 16 ps). Each pixel was acquired for 30 ms acquisition per-pixel in a dry and clear sky environmental conditions leading to an SBR around $\omega=70$. In all simulations using the mannequin head data, we consider $\boldsymbol{\phi}_l = \boldsymbol{\phi} =  (\alpha^{r},\beta^{r},\alpha^{b},\beta^{b}) = (2, \frac{2}{r^M}, 1, \frac{T}{r^M})$ with $ r^M$ being the average number of signal photons per pixel. These  hyper-parameter values are relatively non-informative for both the reflectively and the background noise parameters.

The second scene is the Lego data depicted in Fig. \ref{fig:IRF} and bottom-left of Fig. \ref{fig:MF_GT} (the reader is advised to consult \cite{tobin2017comparative} for more details). The object, of size 42 mm tall and 30 mm wide, was scanned at a standoff distance of 1.8 m using a TCSPC module for an acquisition time per-pixel of 160ms (40ms acquisition time per pixel using four wavelengths where the system IRFs $g_l(.), \forall l$ are shown in Fig. \ref{fig:IRF}-bottom). The size of the spatial, spectral and temporal dimension of the single-photon Lego data are, respectively, $N=200 \times 200$ pixels, $L=4$ wavelengths and $ T = 1500 $ bins with a timing bin size of 2ps. Two versions of the Lego data are used in the experimental section. The first version was acquired in absence of background illumination (SBR$=\omega=66$) and the second one was acquired with presence of ambient illumination leading to an SBR of $\omega=1.3$. In all simulations, the Lego exhibits three classes of interest ($K=3$) whose spectral signatures (related to $\boldsymbol{\alpha^{r}}$ and $\boldsymbol{\beta^{r}}$) are extracted from pixels acquired considering a negligible background contribution and after maximum acquisition time per-pixel (see signatures in Fig. \ref{fig:IRF}). $\alpha_{l}^{b}$ and $\beta_{l}^{b}$ are relatively non-informative such that ($\alpha_{l}^{b},\beta_{l}^{b})=(1, \frac{T}{r_{l}^M})$ with $ r^M_{l}$ being the average number of signal photons per pixel for the $l$th spectral wavelength.
\vspace*{-0.4cm}

{ 
\begin{figure}[h]
\hspace*{-0cm}  
\vspace*{-0cm}


\vspace*{+0cm}
\hspace*{-0.0cm}  
\includegraphics[width=9.5cm]{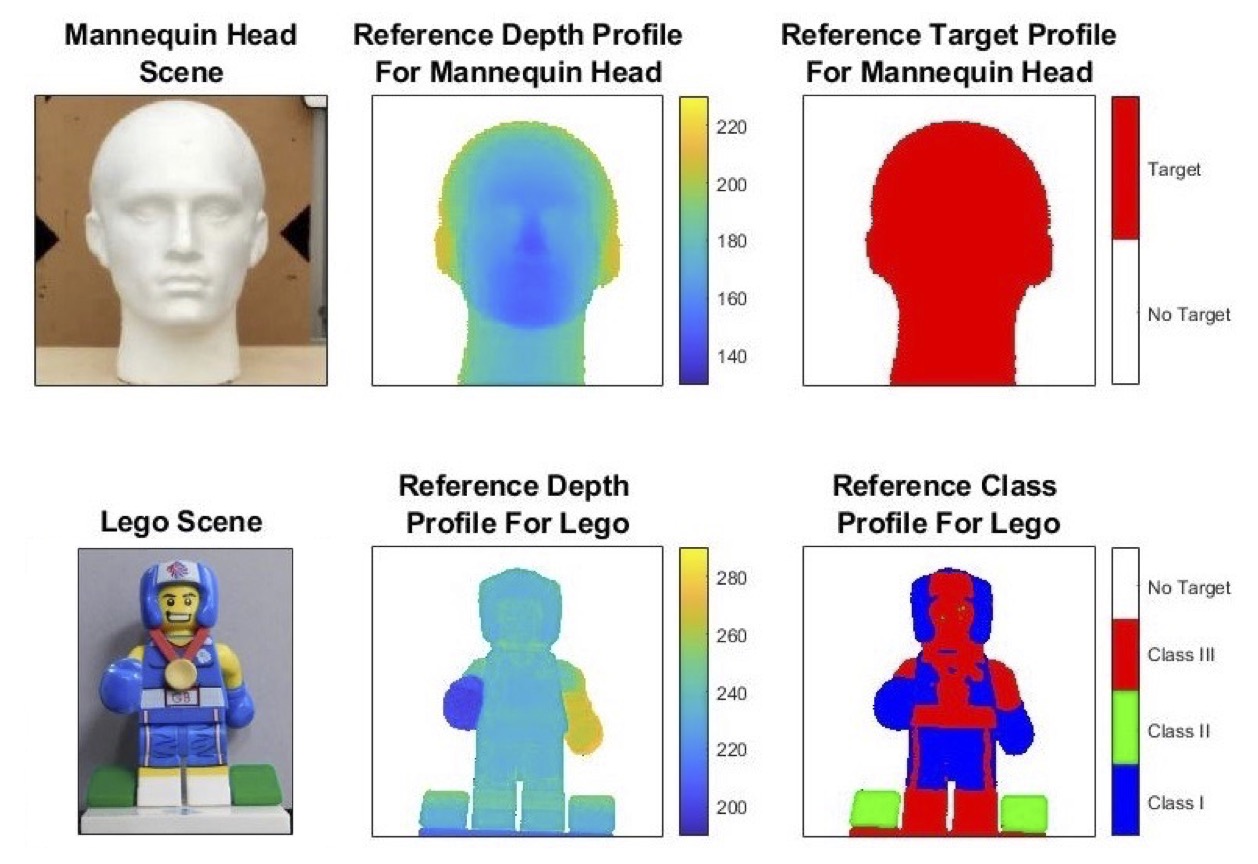}
\caption{From top to bottom: The Mannequin head and the Lego scene. From left to right column: the scene of interest, reference depth estimation (in time bins) and class detection map with maximum acquisition time per-pixel in absence of illumination.}
\label{fig:MF_GT}
\end{figure}
}
\vspace*{-0.0cm}
\subsection{Evaluation of the classification algorithm} \label{subsec:Evaluation_of_the classification_algorithm}

   

\begin{figure}[H]
\centering
\vspace{+0cm}
\includegraphics[width=1.1\figwidth,height=6.5cm]{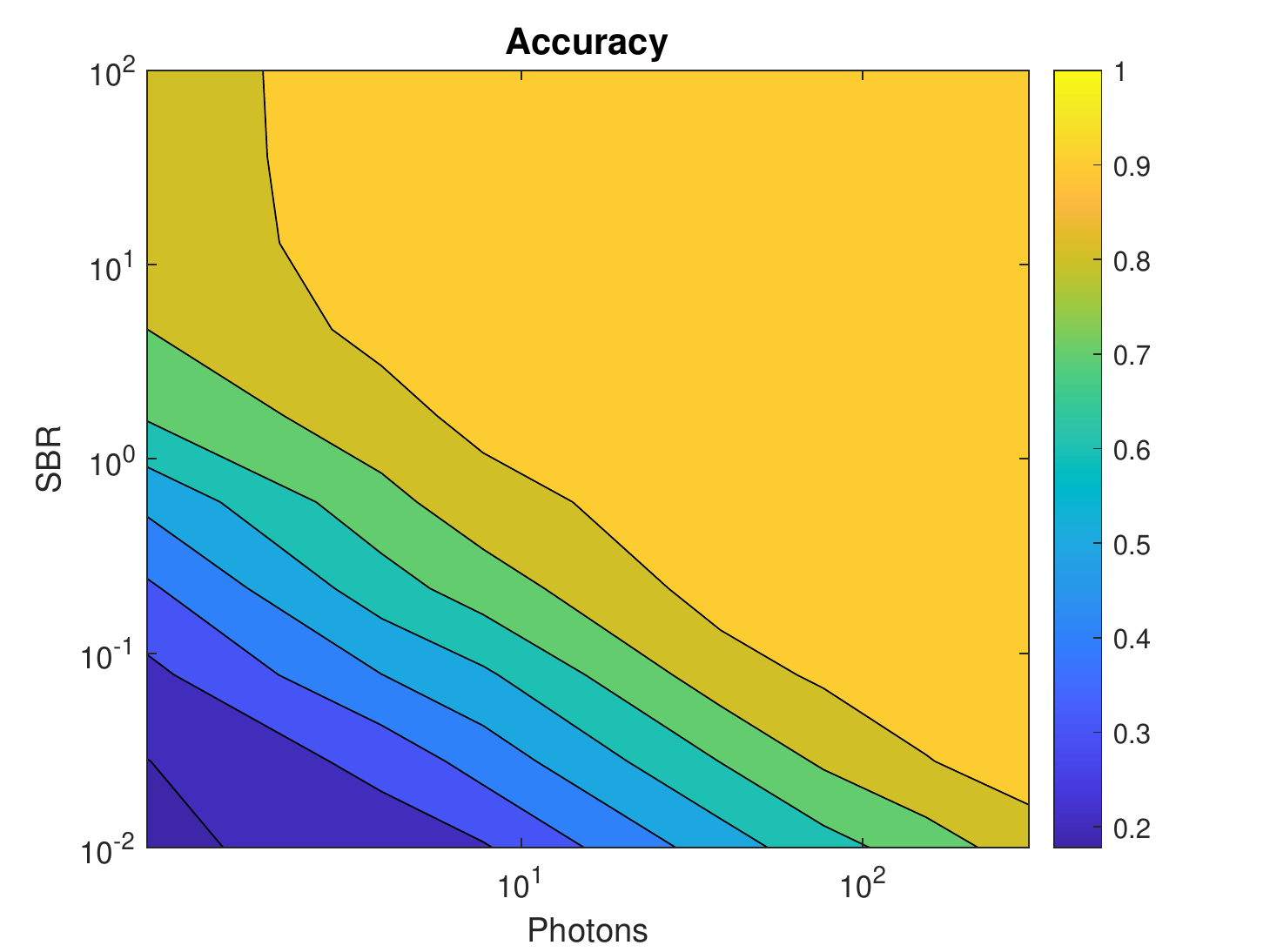}
\caption{(Top) Depth RMSEs and (bottom) Classification accuracy of the sub-sampled Lego data w.r.t. signal-to-background ratio (SBR) and the average signal photons.} 
\label{fig:Accuracy_lego_40by40}
\end{figure}

In this section, we will evaluate the performance of the classification algorithm (according to metrics described in \ref{subsec:Evaluation_criteria}) by simulating data based on the real multispectral single photon Lego data described in Section \ref{subsec:Datasets}. 
We consider a spatially sub-sampled data to analyse the behavior of the algorithm w.r.t. SBR and photons levels.  The subsampled data has $N=40 \times 40$ pixels, $L=4$ wavelengths and $ T = 1500 $ time bins (bin width of 2ps), and is corrupted so that the SBR varies from 0.01 to 100.   Fig. \ref{fig:Accuracy_lego_40by40} top and bottom represent, respectively, the RMSE in meters (top) and the class accuracy (down) w.r.t SBR and the average signal photons. These two figures provide the user with the required number of useful photons (which is proportional to the scanning time) needed to have a given depth precision and accuracy for different SBR levels. Table \ref{tab:confMat_lego_40by40} depicts the confusion matrix for an SBR of magnitude $\omega=0.6$ and after 5 ms of acquisition time per-pixel (which corresponds to signal photons per-pixel $\approx$ 42). In this table, the bold values exhibit the number of pixels of each predicted class,  and bellow them their percentages w.r.t the total number of pixels (here $40^2$ pixels). The last horizontal and vertical lines of this table represent the precision and the recall, respectively. This table highlights a good classification accuracy for these SBR and photon levels. 
Considering previous parameters and for maximum acquisition time, the average computational time per-pixel of the proposed classification algorithm is $\approx$ 55ms, while it should be noted that per-pixel operations are independent allowing parallel processing. 
\begin{figure}[H]
\centering
\includegraphics[width=1.1\figwidth,height=6.5cm]{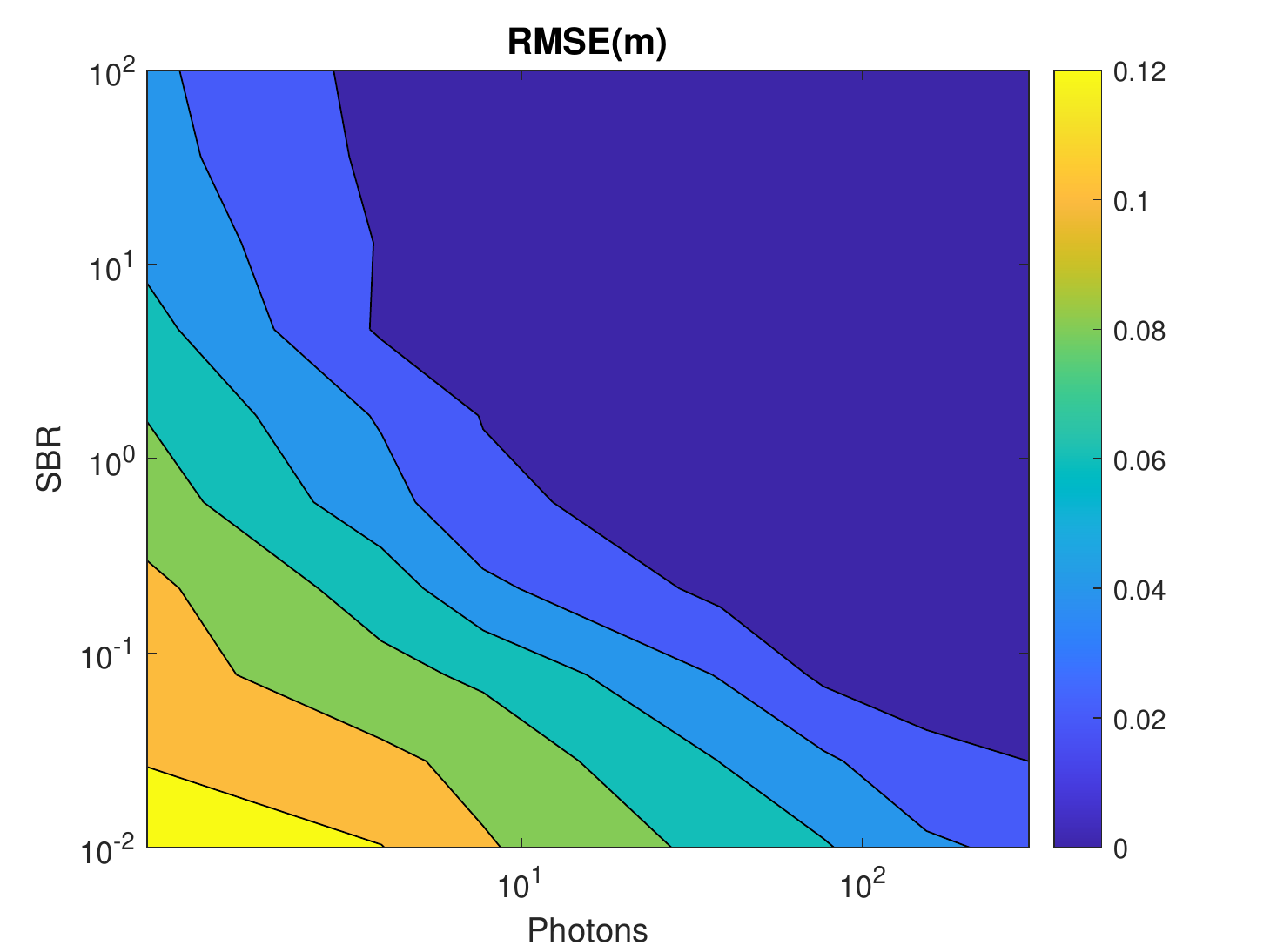}
\vspace{-0.0cm}
\label{fig:RMSE_lego_40by40}
\end{figure} 
\vspace{-2cm}
\vspace{-0cm}
\begin{table}[H] 
\centering
\begin{tabular}{lllll|l}
\cline{2-6} 
\multicolumn{1}{c|}{}  & \multicolumn{4}{c|}{Predicted classes} & \multicolumn{1}{c|}{Recall} \\ 
\hline
\multicolumn{1}{|c|}{} & \multicolumn{1}{c|}{\textbf{223}}  & \multicolumn{1}{c|}{\textbf{9}}  & \multicolumn{1}{c|}{\textbf{6}}  & \multicolumn{1}{c|}{\textbf{1}}  & \multicolumn{1}{c|}{\multirow{2}{*}{93.3$\%$} }  \\ 
\multicolumn{1}{|c|}{True classes}  & \multicolumn{1}{c|}{13.9$\%$}  & \multicolumn{1}{c|}{0.6$\%$}  & \multicolumn{1}{c|}{0.38$\%$}  & \multicolumn{1}{c|}{0.06$\%$} & \multicolumn{1}{c|}{}\\
\cline{2-6} 
\multicolumn{1}{|c|}{} & \multicolumn{1}{c|}{\textbf{0}}  & \multicolumn{1}{c|}{\textbf{63}}  & \multicolumn{1}{c|}{\textbf{1}}  & \multicolumn{1}{c|}{\textbf{2}}      & \multicolumn{1}{c|}{\multirow{2}{*}{95.5$\%$} }  \\
 \multicolumn{1}{|c|}{for}  & \multicolumn{1}{c|}{0$\%$}  & \multicolumn{1}{c|}{3.94$\%$}  & \multicolumn{1}{c|}{0.06$\%$}  & \multicolumn{1}{c|}{0.13$\%$} & \multicolumn{1}{c|}{} \\
\cline{2-6} 
\multicolumn{1}{|c|}{} & \multicolumn{1}{c|}{\textbf{7}}  & \multicolumn{1}{c|}{\textbf{7}}  & \multicolumn{1}{c|}{\textbf{238}}  & \multicolumn{1}{c|}{\textbf{6}}        & \multicolumn{1}{c|}{\multirow{2}{*}{92.3$\%$} } \\
 \multicolumn{1}{|c|}{$\omega = 0.6$}  & \multicolumn{1}{c|}{0.44$\%$}  & \multicolumn{1}{c|}{0.44$\%$}  & \multicolumn{1}{c|}{14.88$\%$}  & \multicolumn{1}{c|}{0.4$\%$} & \multicolumn{1}{c|}{} \\
\cline{2-6} 
\multicolumn{1}{|c|}{} & \multicolumn{1}{c|}{\textbf{0}}  & \multicolumn{1}{c|}{\textbf{10}}  & \multicolumn{1}{c|}{\textbf{0}}  & \textbf{1027}                     & \multicolumn{1}{c|}{\multirow{2}{*}{99$\%$} }  \\
 \multicolumn{1}{|c|}{}  & \multicolumn{1}{c|}{0$\%$}  & \multicolumn{1}{c|}{0.63$\%$}  & \multicolumn{1}{c|}{0$\%$}  & \multicolumn{1}{c|}{64.19$\%$} & \multicolumn{1}{c|}{} \\
\hline
\multicolumn{1}{|c|}{\multirow{2}{*}{Precision} } &
\multicolumn{1}{c|}{\multirow{2}{*}{97.4$\%$} } & \multicolumn{1}{c|}{\multirow{2}{*}{98.1$\%$} } & \multicolumn{1}{c|}{\multirow{2}{*}{83.1$\%$} } & \multicolumn{1}{c|}{\multirow{2}{*}{82.3$\%$} }                    & \multicolumn{1}{c|}{\multirow{2}{*}{96.9$\%$} } \\
\multicolumn{1}{|c|}{}  & \multicolumn{1}{c|}{}  & \multicolumn{1}{c|}{}  & \multicolumn{1}{c|}{}  & \multicolumn{1}{c|}{} & \multicolumn{1}{c|}{} \\
\cline{1-6} 
\label{tab:confMat_lego_40by40}
\end{tabular}
\caption{{Confusion matrix of the sub-sampled Lego data ($40 \times 40$ pixels) for $\textrm{SBR}= 0.6$  and $5$ milliseconds of dwell time per-pixel and per wavelength (signal average photons per-pixel $\approx$ 42).}}
\end{table}

 \vspace{-0cm}





\begin{figure*}
\centering
\subfigure[Sequential acquisition]{%
\includegraphics[width=0.60\figwidth,height=5.0cm]{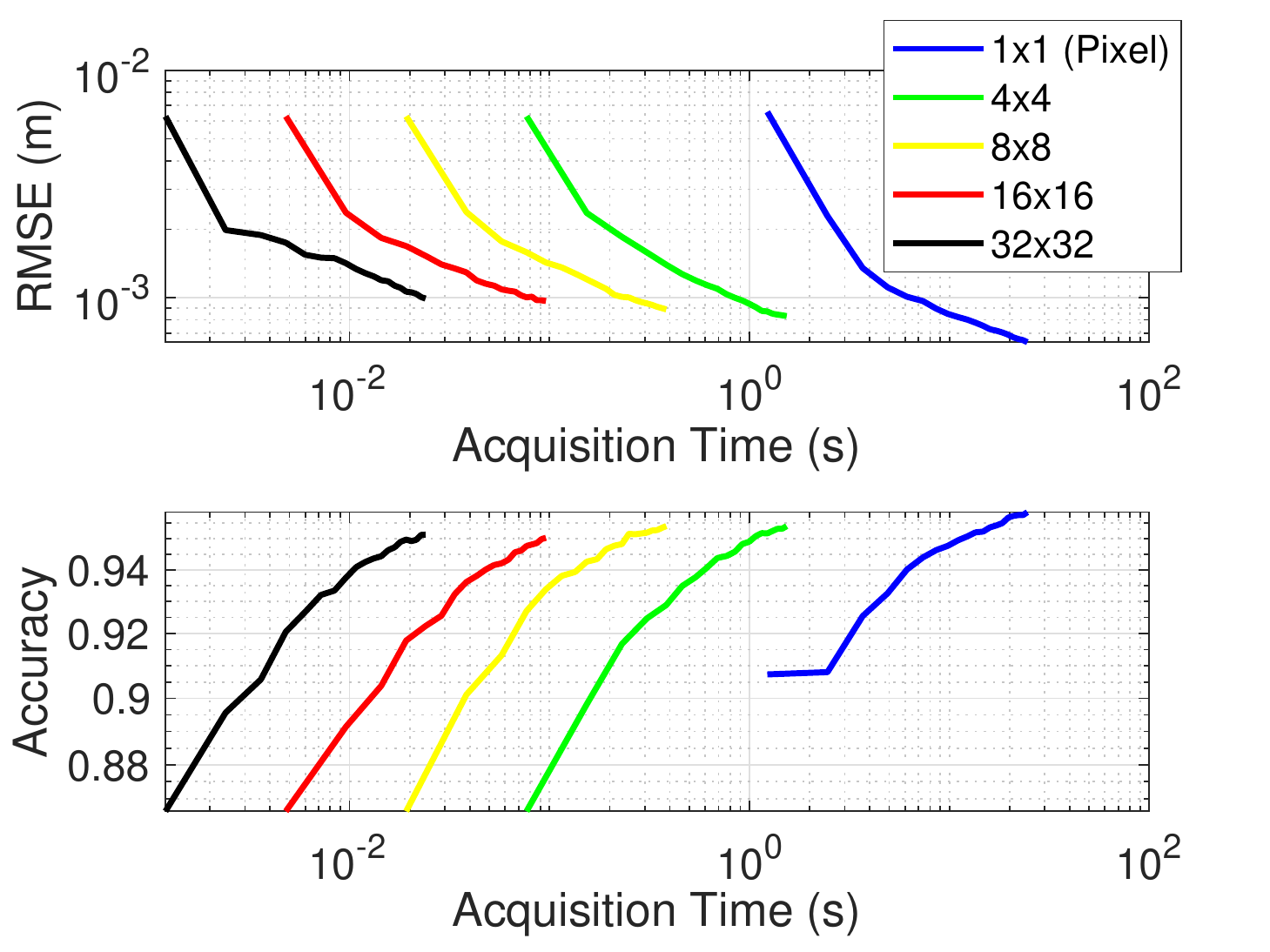}
\label{fig:Sequential_ACC_RMSE_HSBR_K3}%
}\qquad
\subfigure[Parallel acquisition]{%
\includegraphics[width=0.60\figwidth,height=5.0cm]{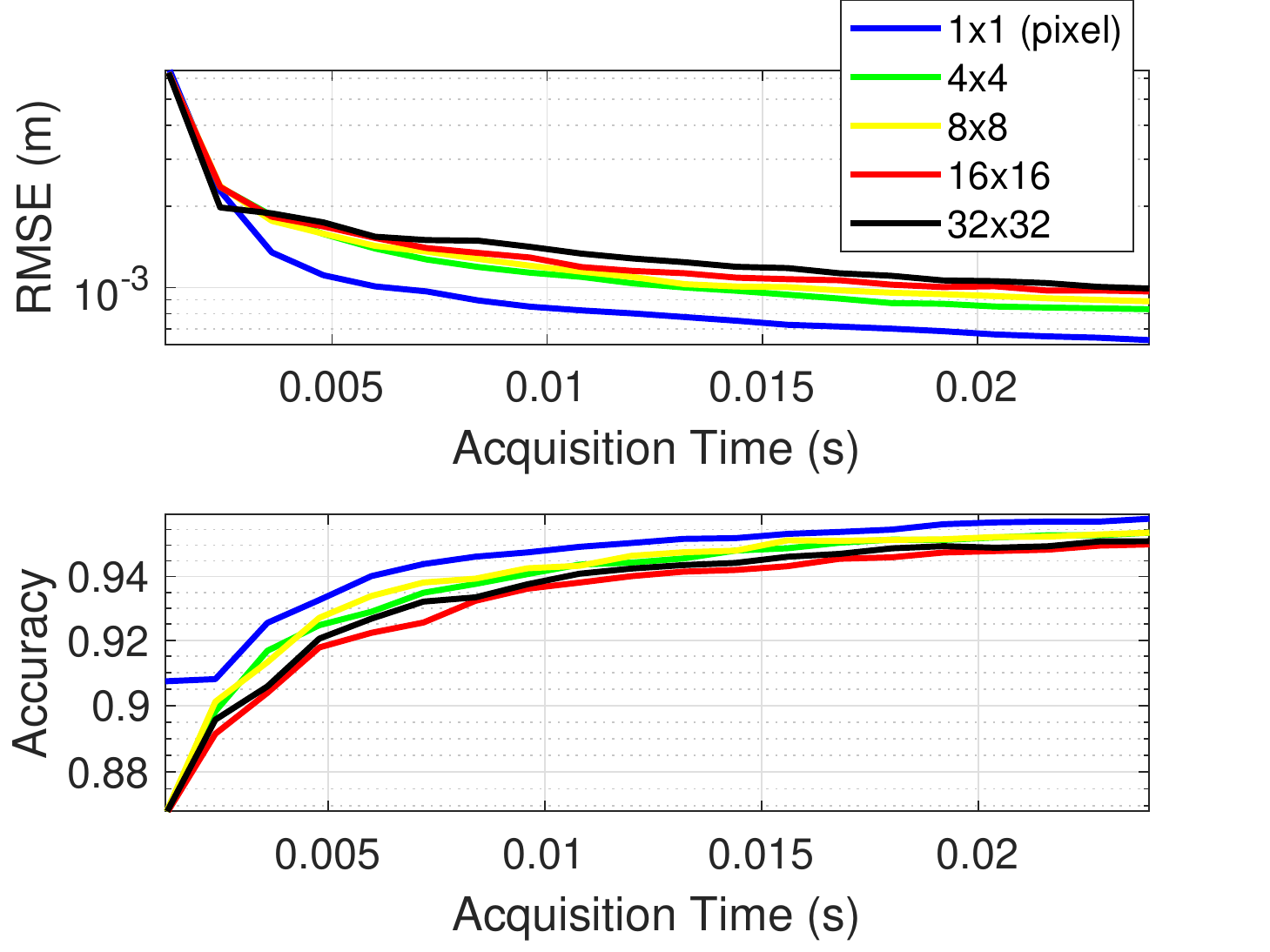}
\label{fig:Parallel_ACC_RMSE_HSBR_K3}%
}
\subfigure[Sampling pattern]{%
\includegraphics[width=0.60\figwidth,height=5.0cm]{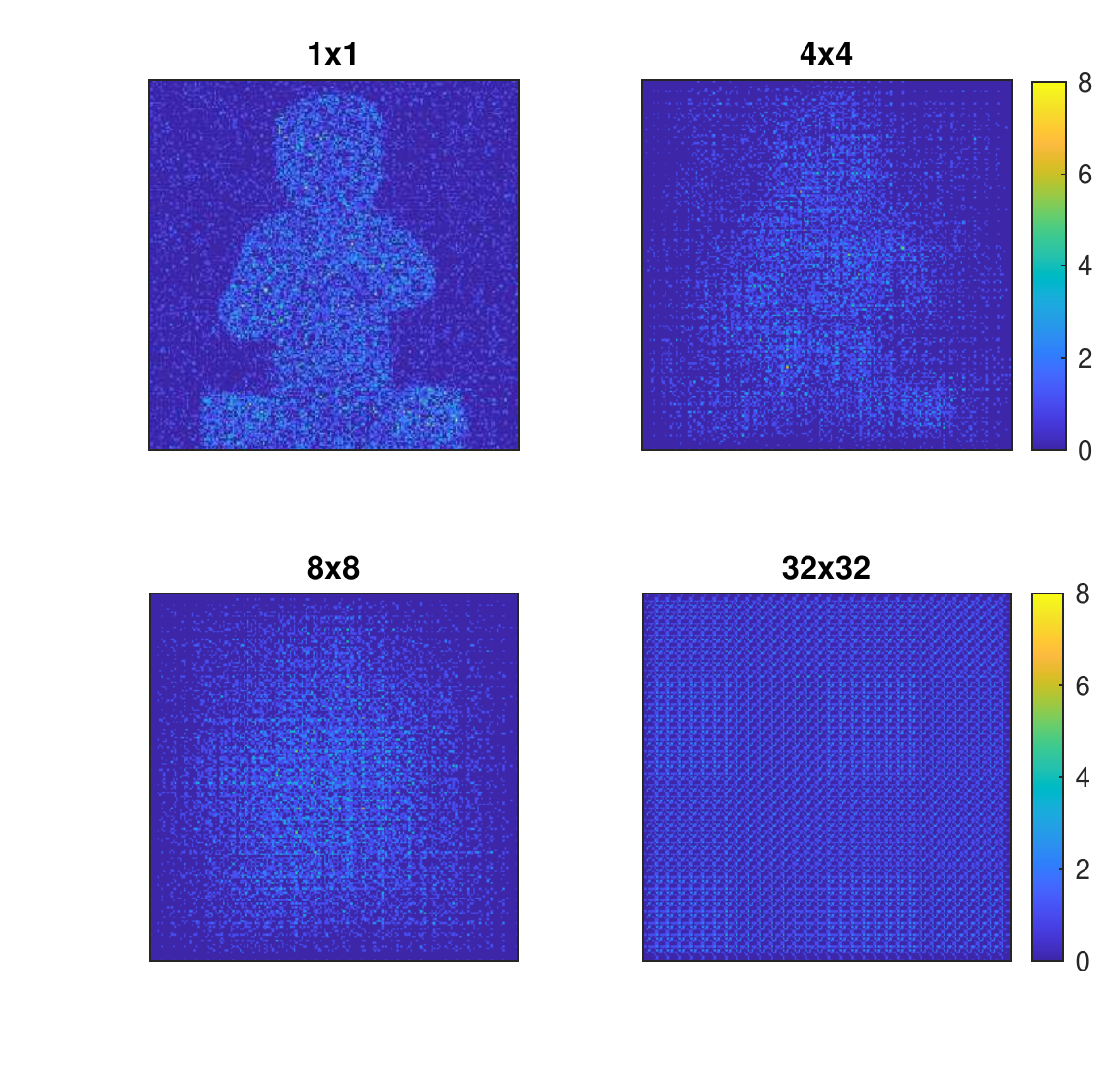}
\label{fig:Parsec_AC_pattern_HSBR_K3}%
}\qquad

\caption{Depth RMSEs and classification accuracy w.r.t.  acquisition times for different SPAD array resolution when the acquisition mode is done (a) sequentially, or (b) in parallel. (c) Depicts the sampling pattern for different array sizes.}
\label{fig:sequential_parallel_pattern}
\end{figure*} 

\vspace{0cm}
\subsection{Pixel and array based AS} \label{subsec:Pixel_and_array_based_AS} 

This section analyses the performance of the proposed AS strategy when considering different scanning scenarios as indicated in Fig. \ref{fig:Scan_scenarios}. More precisely, we study sequential and parallel sampling, when considering different array sizes. For all cases, we evaluate the depth and class accuracy with respect to acquisition time. As indicated in Section \ref{subsec:sample_generation},  the acquisition time per one iteration of AS will depend on the considered sampling scenario (e.g., equal to $N_s t_0/r^2 $ for sequential scanning and to $t_0$ for parallel scanning, when assuming $t_0$ dwell time per shot and $r \times r$ array). 
The parameters considered for this section are $N_s=32^2$, $t_0=300 \mu s$. 


First, we fix  SBR$= 66$ and study the effect of varying array sizes as follows $r \in \left\lbrace 1, 4, 8, 16, 32 \right\rbrace$.
Fig. \ref{fig:Sequential_ACC_RMSE_HSBR_K3} and Fig. \ref{fig:Parallel_ACC_RMSE_HSBR_K3} shows the variation of depth RMSE and classification accuracy w.r.t. acquisition time. As expected, these figures show faster convergence for parallel acquisition reaching millisecond levels as opposed to seconds in sequential scanning. Considering sequential scanning,  Fig. \ref{fig:Sequential_ACC_RMSE_HSBR_K3} highlights faster convergence for larger array systems as they acquire $r^2$ samples in parallel. For parallel scanning, large arrays impose spatial constraints on the sampled locations, hence, better performance is obtained by small arrays as they better approximate the scene features. This effect is highlighted in Fig. \ref{fig:Parsec_AC_pattern_HSBR_K3} showing the sampled points, where smaller arrays allow fine scene scanning by locating more points on the Lego shape (i.e., dense and focused scanning pattern) than larger arrays.

\begin{figure*}
\centering
\subfigure[$1\times 1$ pixel scanning]{%
\includegraphics[width=0.63\figwidth,height=5.3cm]{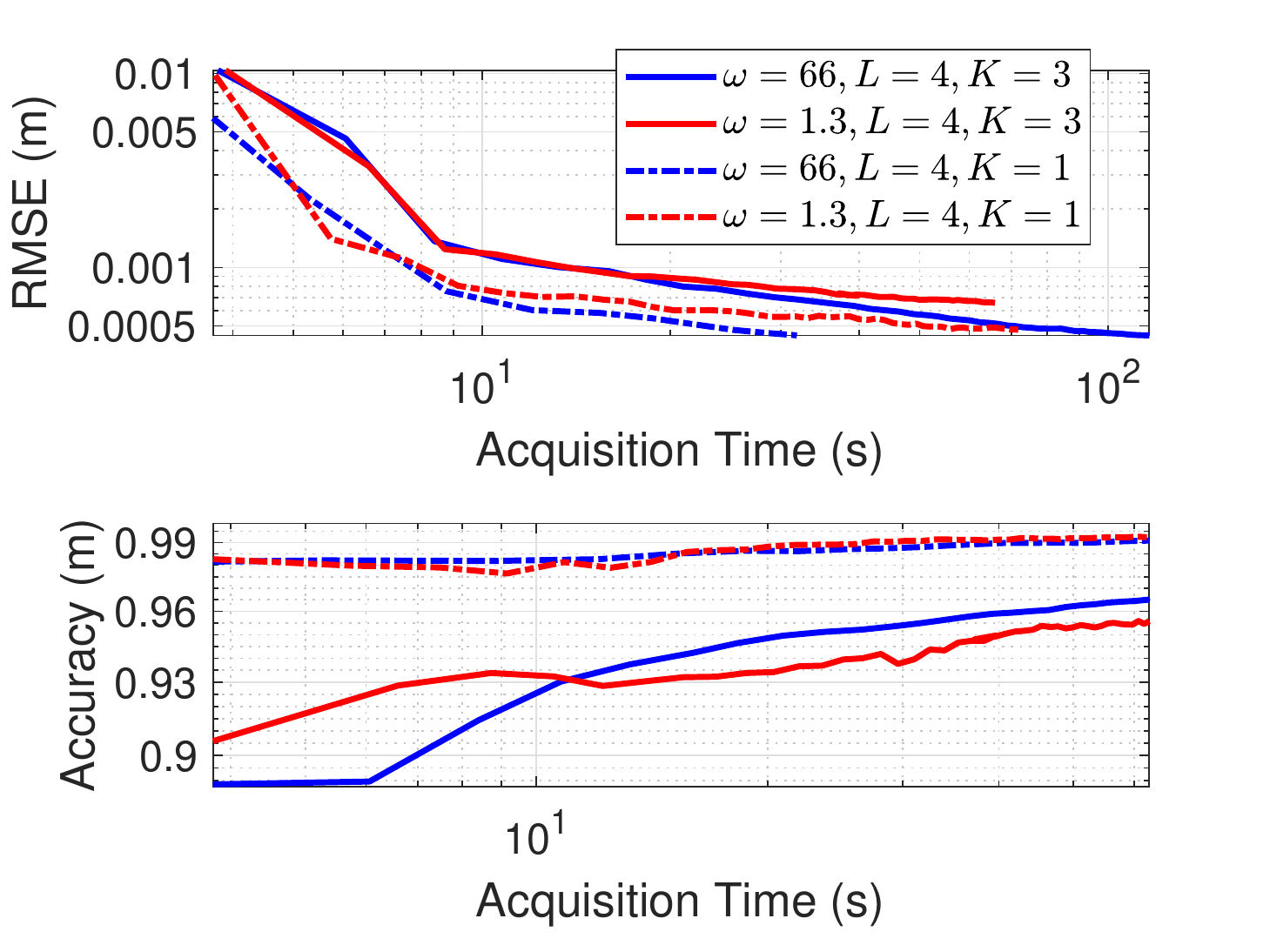}
\label{fig:AS_pixel_ACC-RMSE_SBR_K}%
}\qquad
\subfigure[$32\times 32$ array scanning]{%
\includegraphics[width=0.63\figwidth,height=5.3cm]{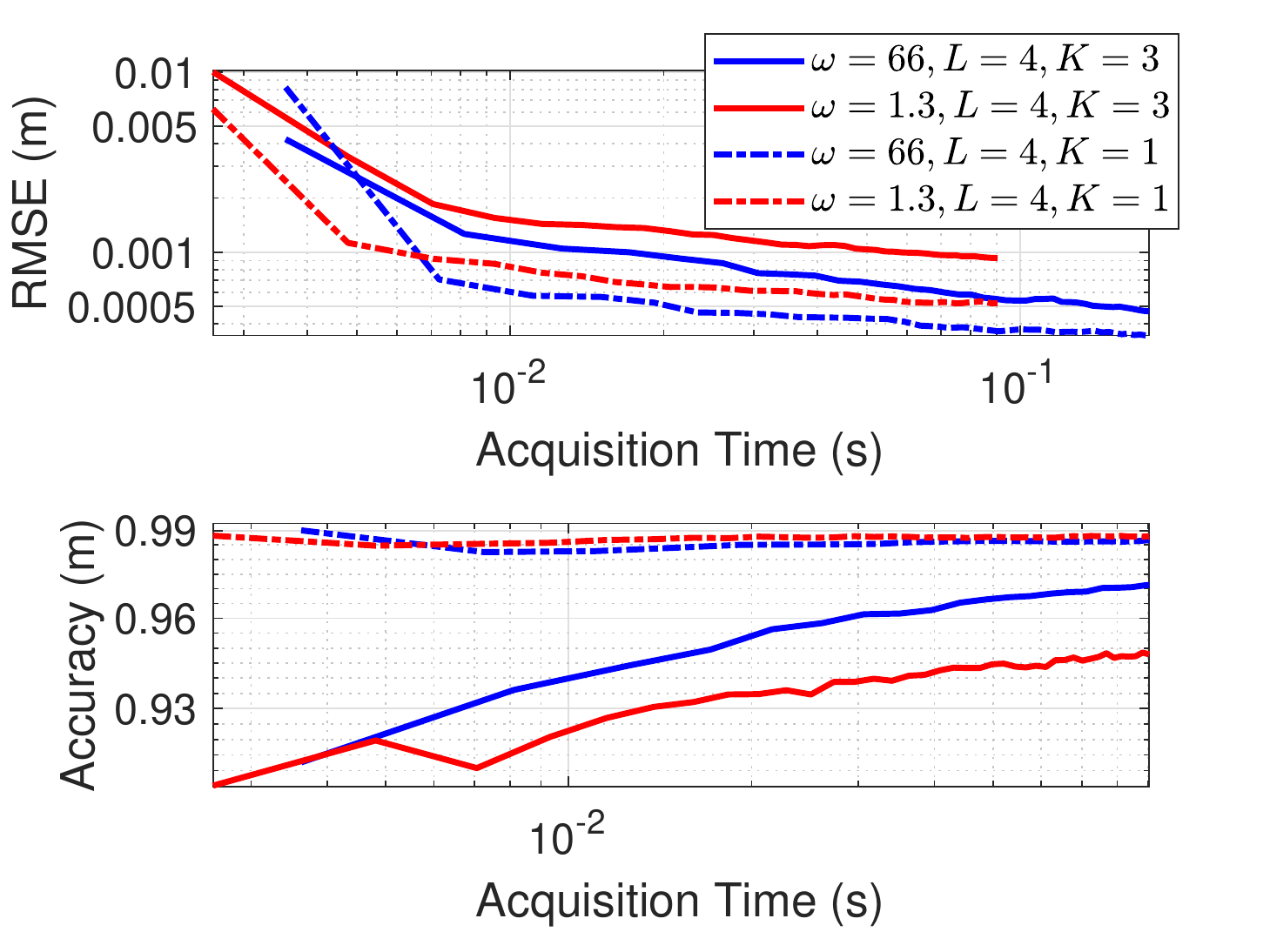}
\label{fig:AS_array_ACC-RMSE_SBR_K}%
}
\subfigure[Scanning Pattern]{%
\includegraphics[width=0.63\figwidth,height=5.3cm]{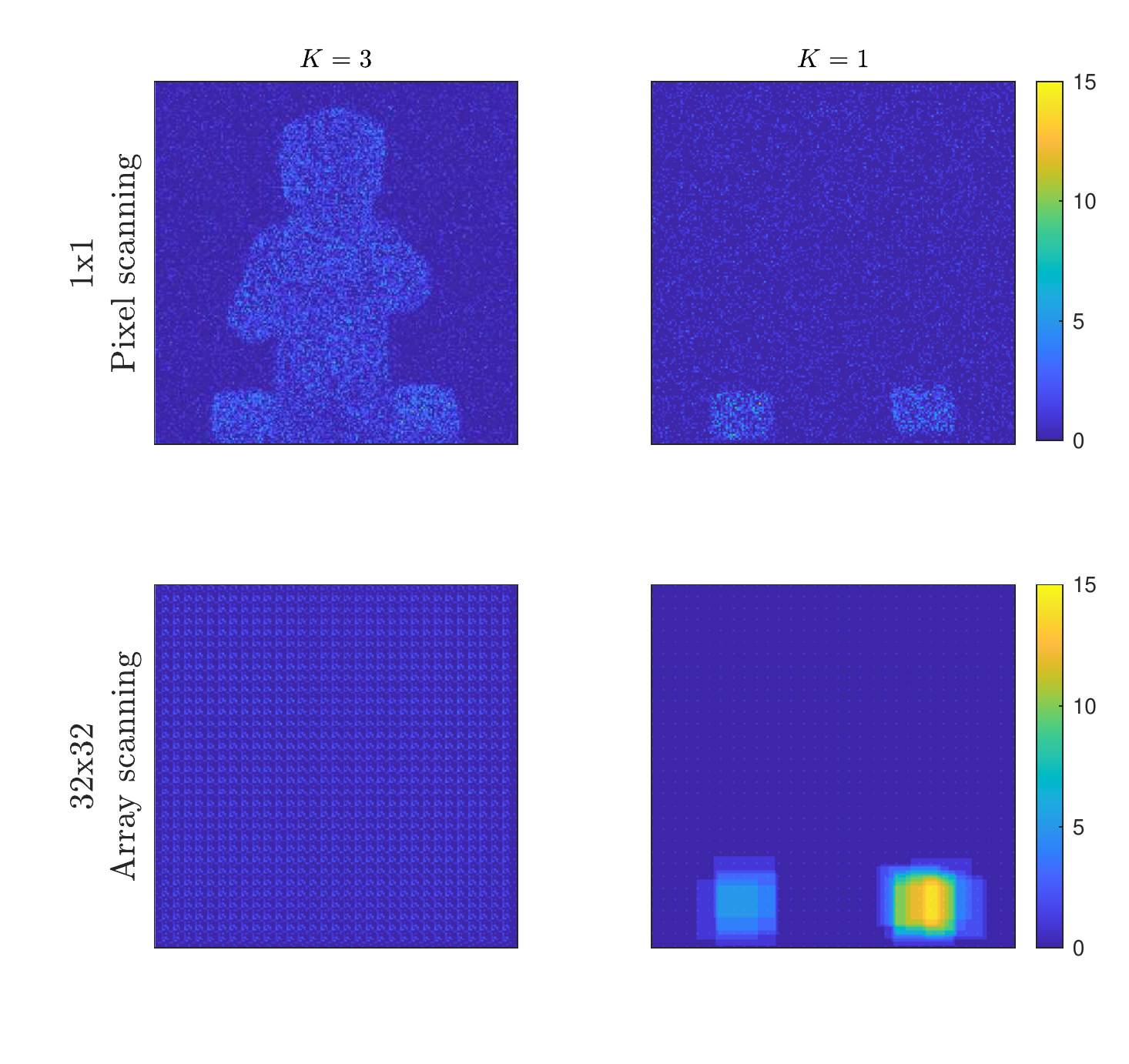}
\label{fig:array_pixel_scan_v2}%
}\qquad
\vspace{-0.0cm}
\caption{Depth RMSEs and classification accuracy w.r.t.  acquisition times for two SBR levels for multispectral target detection ($K=1$) and spectral classification ($K=3$) using a (a) $1 \times 1$ pixel scanning system, (b) $32\times32$ array scanning system. (c) Depicts the sampling pattern obtained for $r \in \left\lbrace 1, 32 \right\rbrace$ when considering the full Lego as a target (i.e., $K=3$) or only the green region (i.e., $K=1$).}
\label{fig:pixel_array_pattern}
\end{figure*}

Second, we evaluate the effect of SBR and the size of the object of interest within the field of view,
when considering a sequential acquisition and $r \in \left\lbrace 1, 32 \right\rbrace$. To change the target size,  we compare a large ROI obtained when targeting the full Lego (i.e., the $K=3$ classes are the target) and the case where the target of interest are the green pixels in Fig. \ref{fig:IRF} (i.e., the target is only one class $K=1$). 
Fig. \ref{fig:AS_pixel_ACC-RMSE_SBR_K} and Fig. \ref{fig:AS_array_ACC-RMSE_SBR_K} show depth and accuracy performance for different SBR$\in \left\lbrace 1.3, 66 \right\rbrace$ and object sizes. 
As expected, the proposed AS approach shows faster convergence for smaller regions of interest (i.e., dashed lines better than continuous lines), and/or for larger SBR values (i.e., blue curves are similar or slightly better than red ones). Fig. \ref{fig:array_pixel_scan_v2} shows the sampled patterns for the different cases, indicating finer scanning for $r=1$ compared to coarse results  when considering $r=32$. 



\vspace{-0.0cm}
\section{Results on real data} \label{sec:Results_on_real_data} 

This section evaluates the proposed strategy on two experimental data sets, namely the monochromatic mannequin head and the multispectral Lego scene. In both cases, we compare the proposed adaptive sampling strategy against static sampling strategies and highlight its benefits for range estimation, both in absence and presence of ambient illumination cases.  
The considered static sampling strategies include the  uniform  (US) and random (RS) sampling strategies that only scan a ratio of the $N$ pixels (we considered 30$\%$ and 60$\%$ of pixels for random sampling). For these strategies, the maximum likelihood estimate in absence of background counts is considered for the depth \cite{mccarthy2013kilometer}, which reduces to the maximum of the correlation between the log-IRF and each histogram (denoted by Xcorr for cross-correlation). For multispectral data, the depth estimate reduces to maximizing the sum of the cross-correlations between each log-IRF and its corresponding histogram channel-wise, which corresponds to the maximum likelihood estimator in absence of background counts. The median operator $h_p$ is applied to these depth estimates to fill unscanned pixels. Finally, the system that will be used is a pixel-wise SPAD system ($r=1$) and the acquisition mode is sequential.

\subsection{Evaluation of AS on the Mannequin head}
\label{subsec:Evaluation_of_AS_with_target_detection} 

\begin{figure*}[t]
\centering
\subfigure[$\omega=70$]{%
\includegraphics[width=0.9\figwidth,height=4cm]{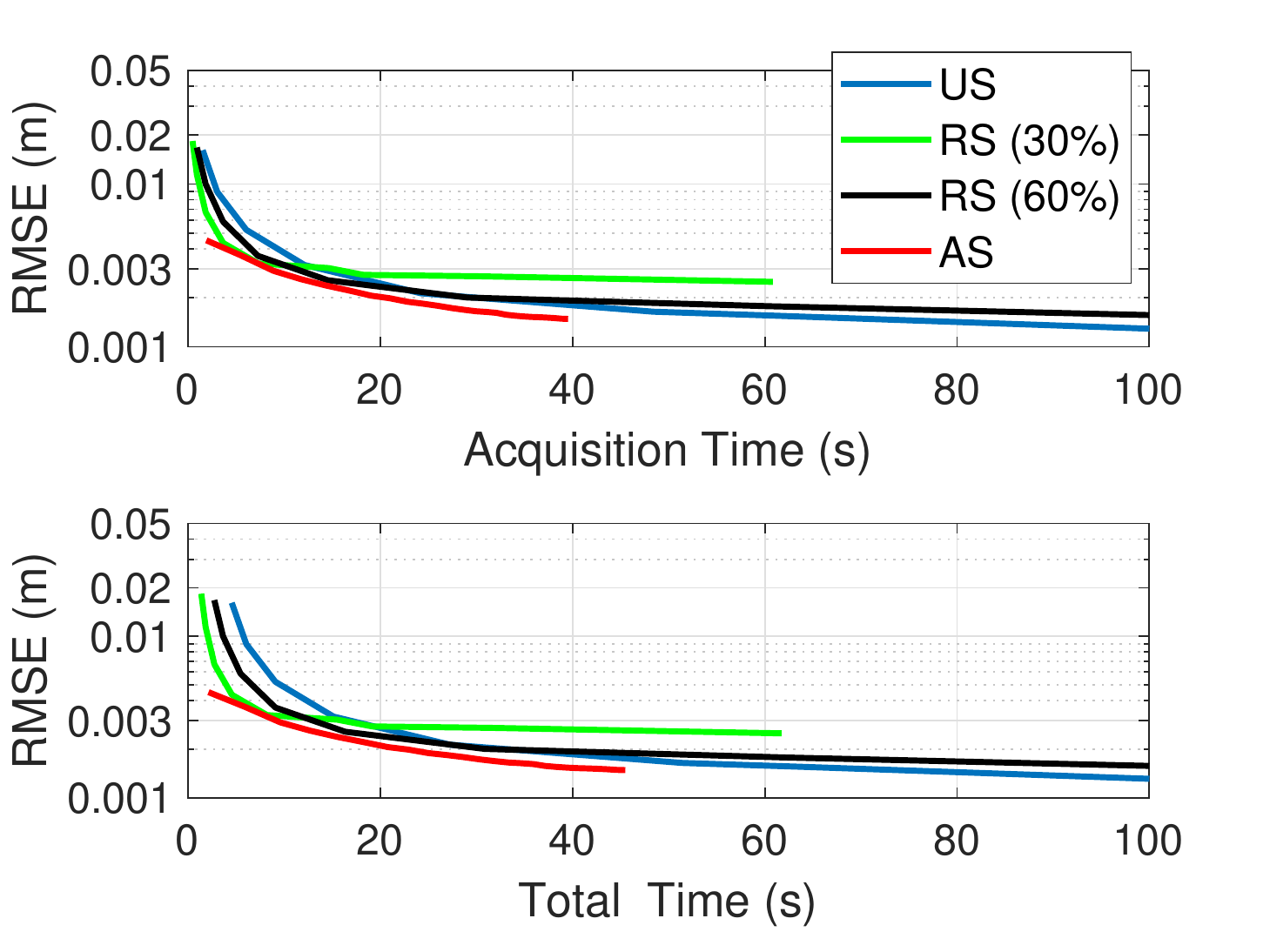}
\label{fig:RMSE_MF_clean}
}\qquad
\subfigure[$\omega=0.48$,]{%
\includegraphics[width=0.9\figwidth,height=4cm]{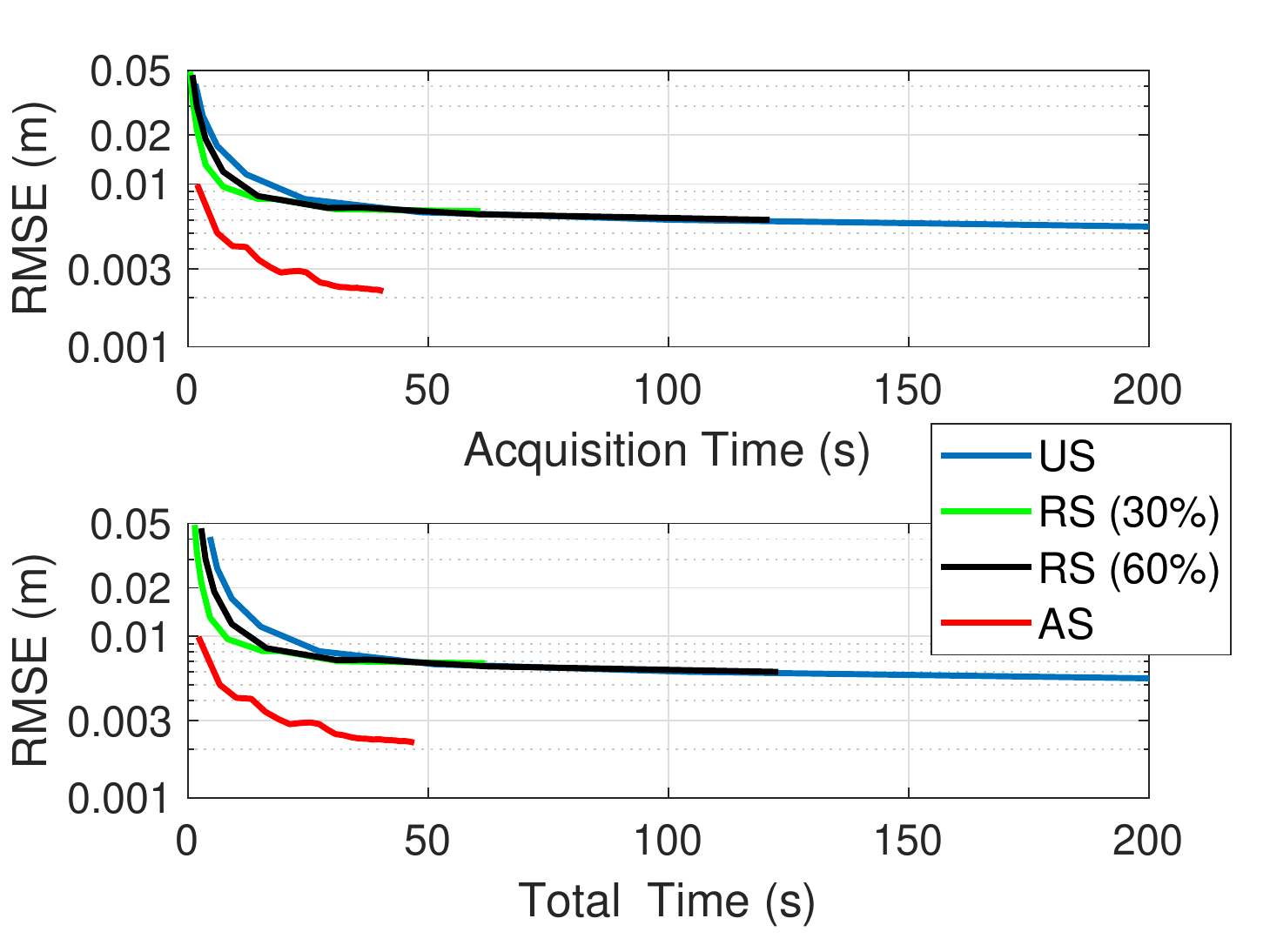}
\label{fig:RMSE_MF_LowSBR}
}\\
\subfigure[$\omega=0.18$]{%
\includegraphics[width=0.9\figwidth,height=4cm]{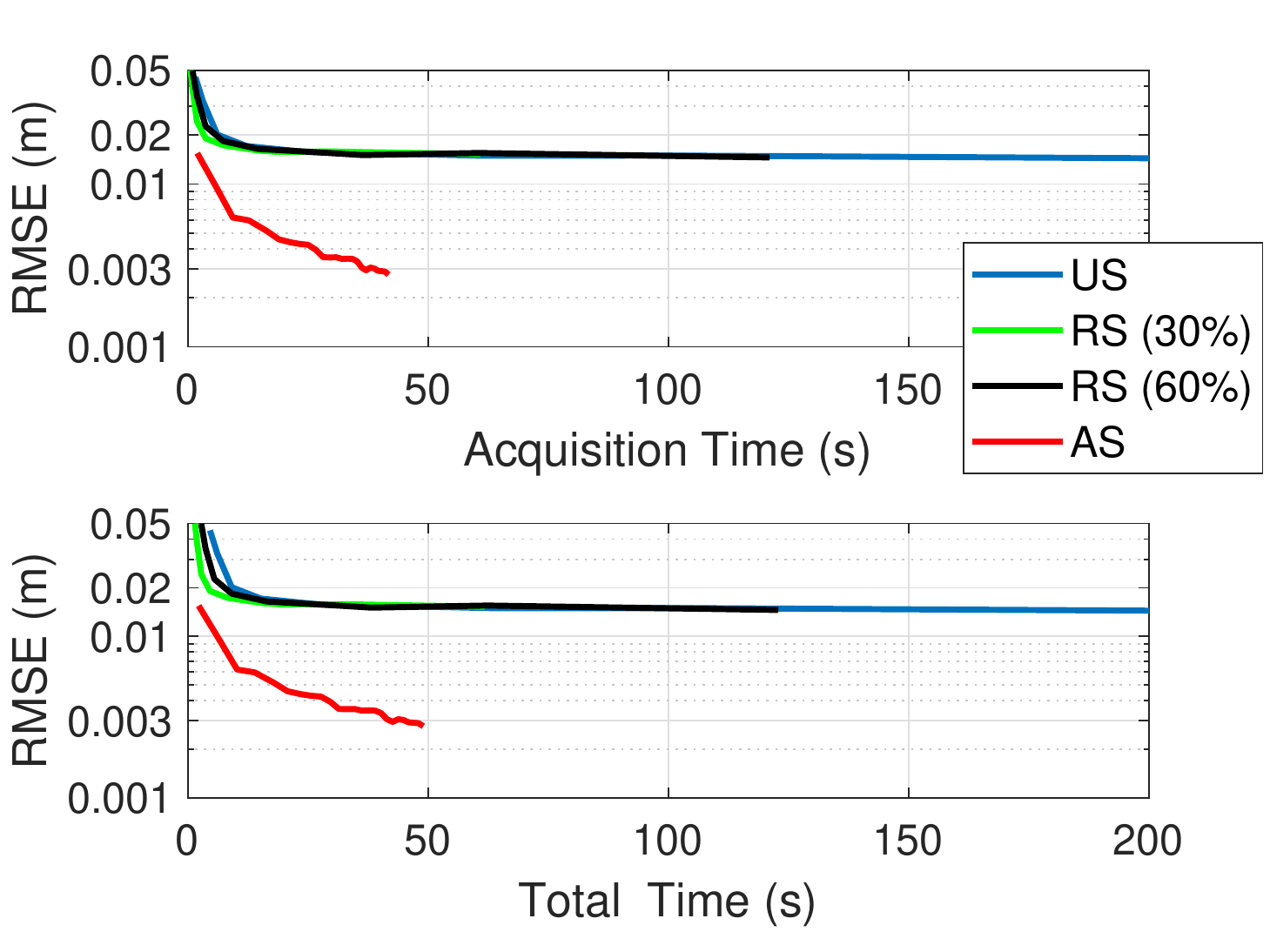}
\label{fig:RMSE_MF_VeryLowSBR}
}\qquad
\subfigure[Target accuracy for different $\omega$]{%
\includegraphics[width=0.9\figwidth,height=4cm]{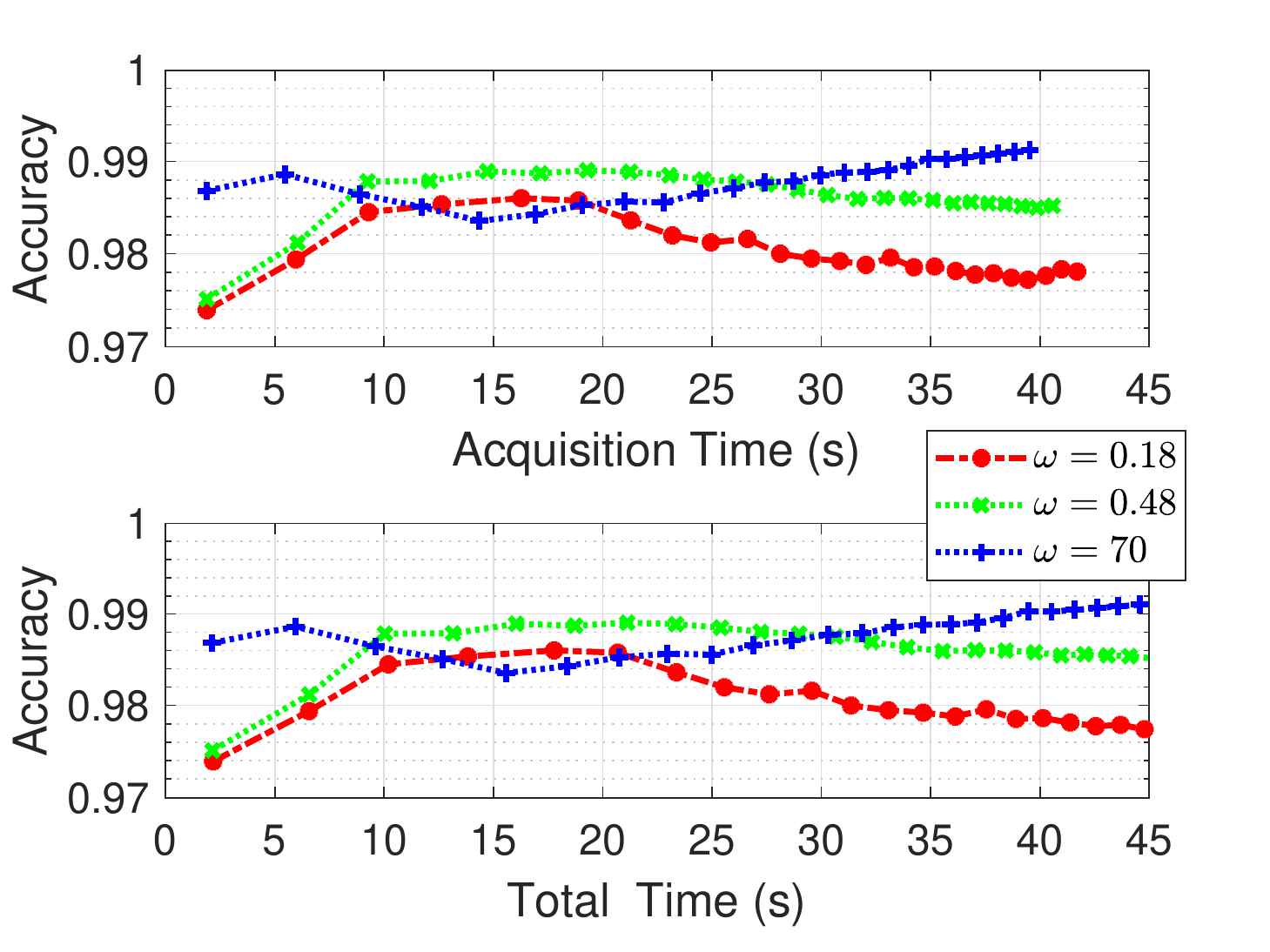}
\label{fig:acc_MF_Bayes_EverySBR}

}
\caption{Depth RMSEs of the full mannequin head target for different sampling strategies w.r.t. (top) dwell time and (bottom) total time  when (a) $\omega=70$, (b) $\omega=0.48$, (c) $\omega=0.18$. (d) Target accuracy  of the full mannequin head target for different sampling strategies w.r.t. (top) dwell time and (bottom) total time for three SBR levels.}
\label{fig:depth_uncer_class}
\end{figure*} 

This subsection compares the proposed AS approach to other static sampling strategies. 
We consider the mannequin head scene and focus on the particular case where the number of wavelength and classes of interest are respectively $L=1$ and $K=1$. This reduces the classification algorithm to a target detection algorithm. 
The proposed algorithm estimates the parameters of interest (depth and class) in a pixel-wise fashion, thus we report pixel-wise values as the processing can be parallelized. 
We compare for two SBR levels, a high SBR= 70 with almost no background light in Figs. \ref{fig:RMSE_MF_clean}, and a noisy case with SBR $=0.48$ and SBR$=0.18$ in Figs. \ref{fig:RMSE_MF_LowSBR} and \ref{fig:RMSE_MF_VeryLowSBR}, respectively. The accuracy of all SBR levels are reported in Figs. \ref{fig:acc_MF_Bayes_EverySBR}. In any of the above figures, the top subplots evaluate performance w.r.t. the acquisition  time (dwell time) used to scan the data. We also evaluate performance while accounting for the total time of the AS process which includes:  the  dwell time, the per-pixel processing time to perform the task, the time to build the ROI map and the time to move the scanning mirrors (one move is approximated by $150 \mu$s). Under high SBR, Figs. \ref{fig:RMSE_MF_clean} show that depth performance are similar for the different sampling strategies, although the proposed algorithm provides additional classification information. The benefit of the proposed framework becomes clear in the noisy case, where an improvement factor of $\simeq$ 100 is observed compared to static sampling strategies using Xcorr (at depth RMSE equal to 15mm in Fig. \ref{fig:RMSE_MF_VeryLowSBR}). Note that target accuracy is higher than $97\%$ at the very beginning of the sampling process regardless of the SBR level thanks to the robust restoration strategy performed by $h_p$, $h_u$ using the median operator.

\begin{figure*}[h]
\centering
\subfigure[$\omega=66$, $K=3$]{%
\includegraphics[width=0.9\figwidth,height=4cm]{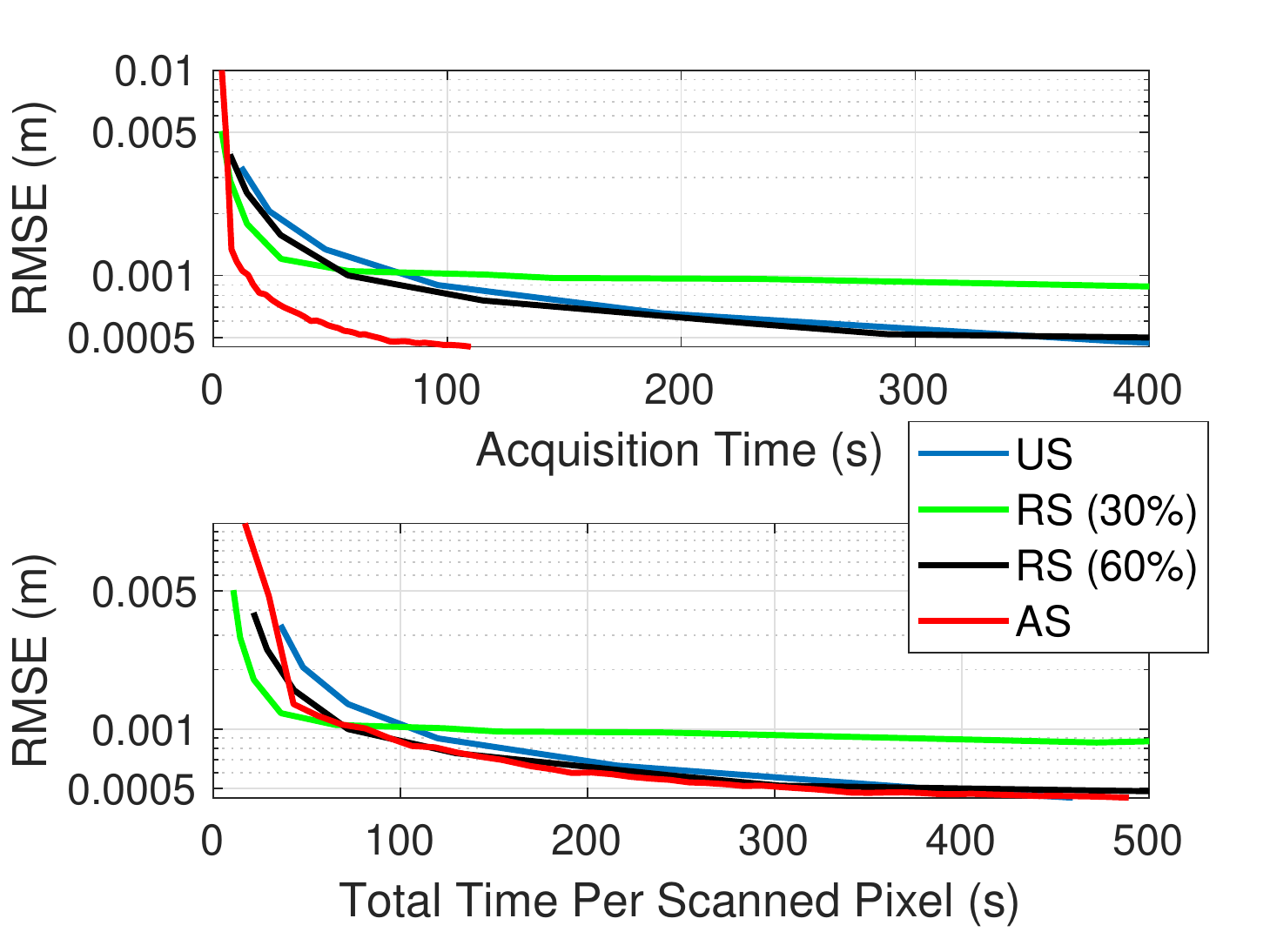}
\label{fig:RMSE_lego_clean_usrsas}%
}\qquad
\subfigure[$\omega=66$, $K=1$]{%
\includegraphics[width=0.9\figwidth,height=4cm]{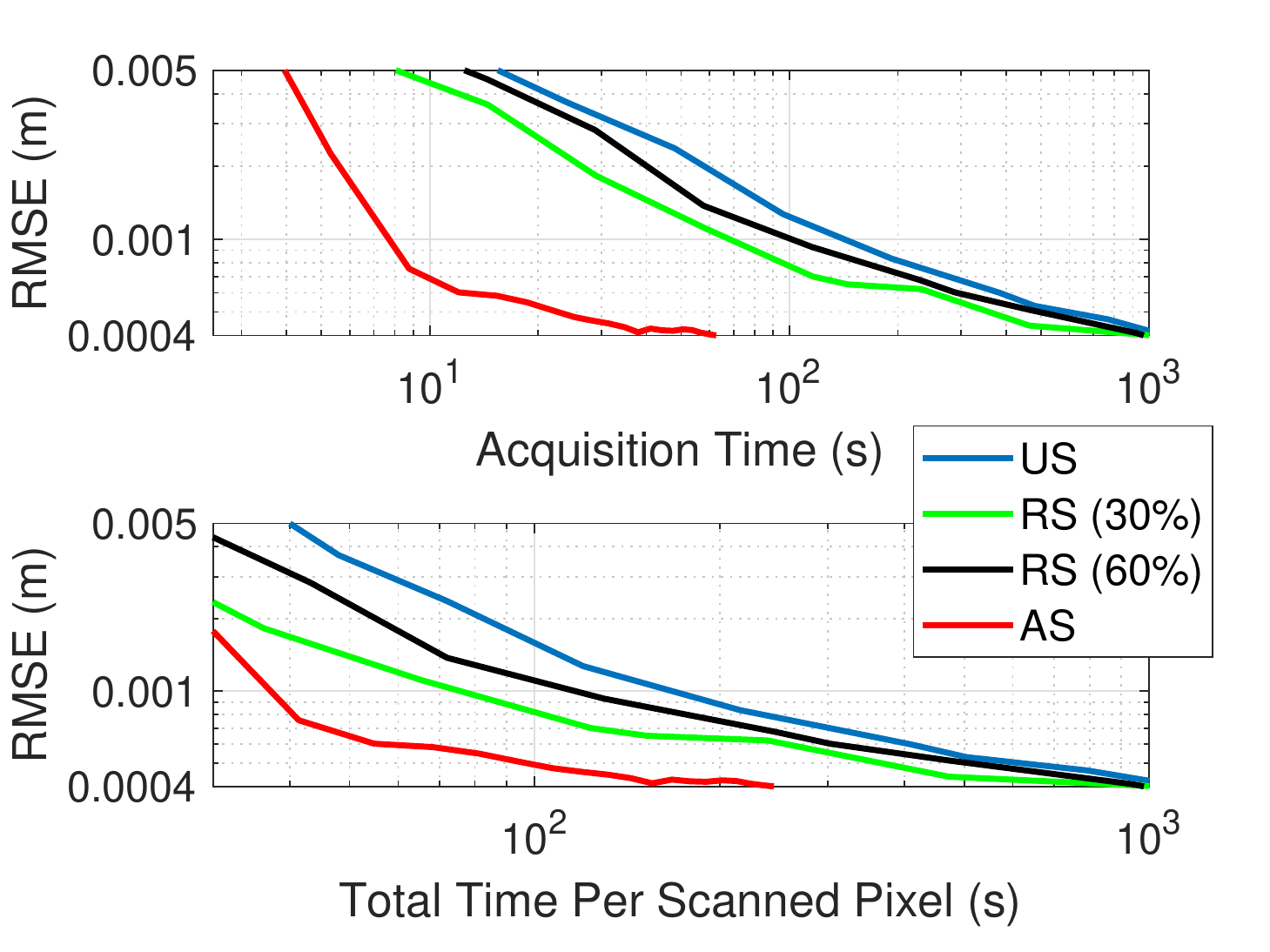}
\label{fig:RMSE_lego_clean_one_class_usrsas}%
}\\
\subfigure[$\omega=1.3$, $K=3$]{%
\includegraphics[width=0.9\figwidth,height=4cm]{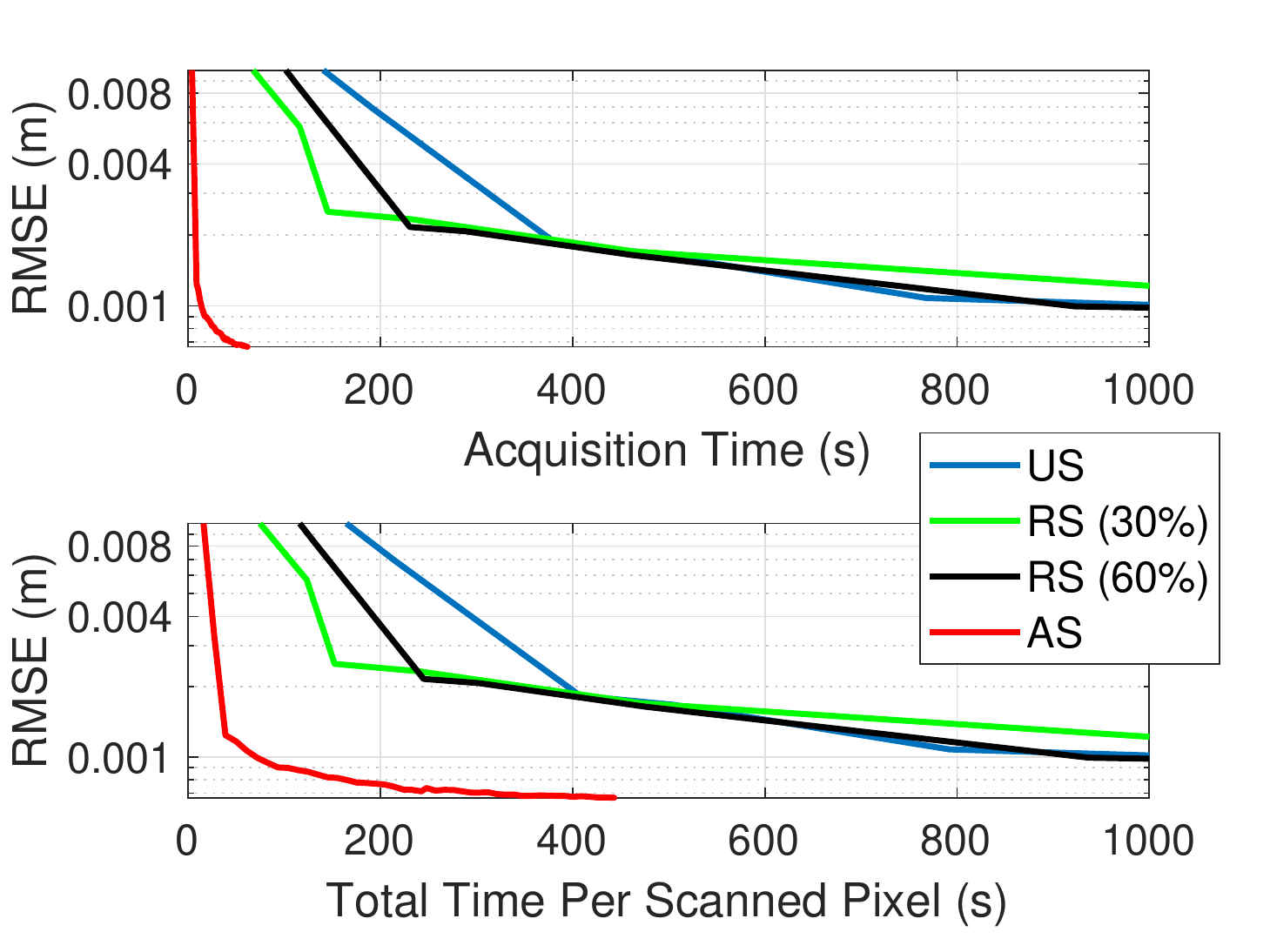}
\label{fig:RMSE_lego_dirty_usrsas}%
}\qquad
\subfigure[$\omega=1.3$, $K=1$]{%
\includegraphics[width=0.9\figwidth,height=4cm]{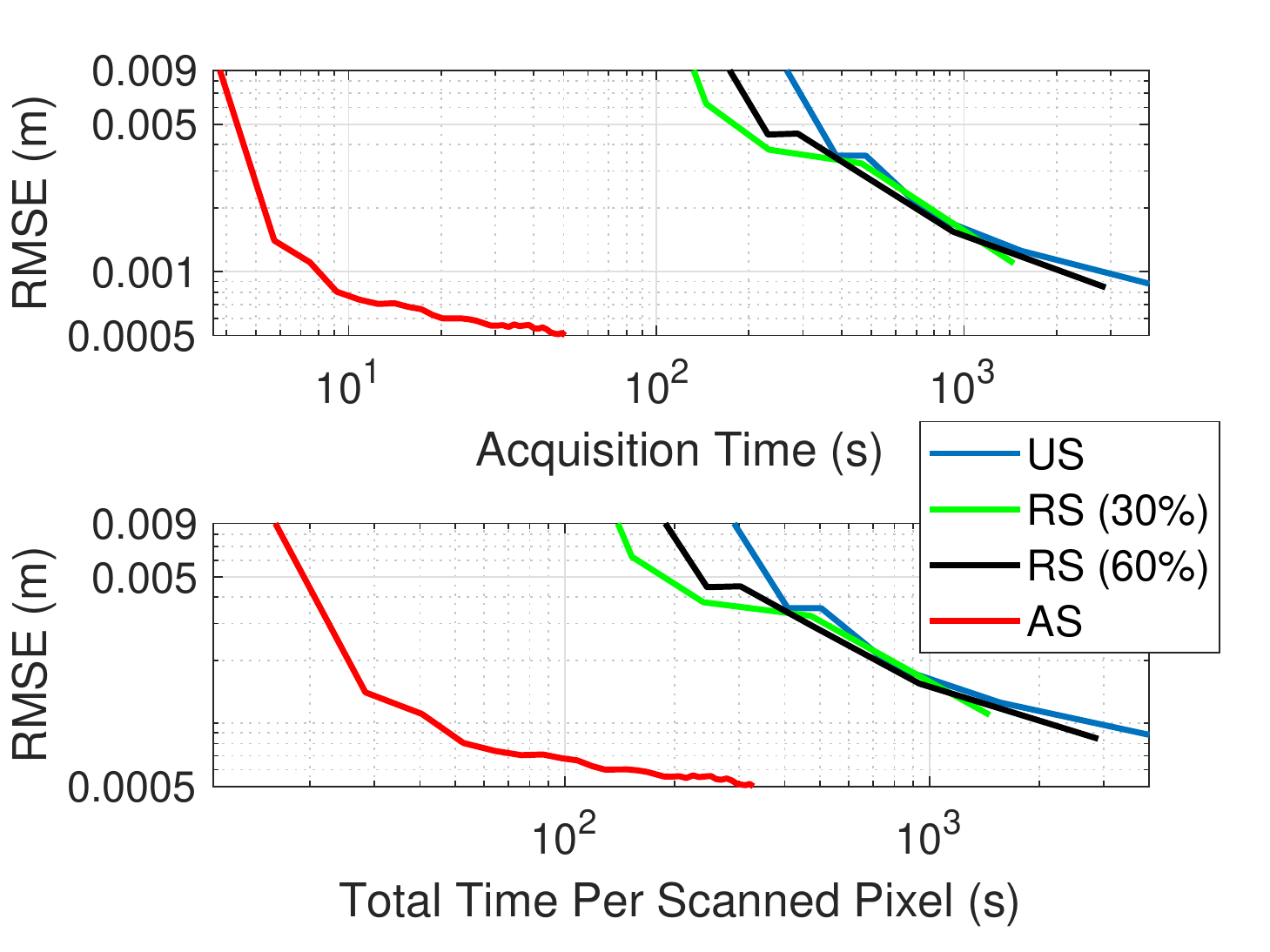}
\label{fig:RMSE_lego_dirty_one_class_usrsas}%
}
\caption{Results of the proposed algorithm w.r.t acquisition time (dwell time) and total time compared to static sampling strategies for number of classes and two SBR levels targeting (a) 3 classes and (b) 1 class in the absence of background illumination, (c) 3 classes and (d) 1 class under the presence of background illumination.}
\label{fig:lego_rmse_as_lowHigh_one_three}
\end{figure*} 




\subsection{Evaluation of AS on the multispectral Lego scene}
This subsection evaluates the proposed strategy on the full multispectral Lego data described in Section \ref{subsec:Datasets} (see  Fig. \ref{fig:IRF} (top)).
The approach is compared to static sampling strategies while considering two tasks: (i) classification based on spectral signature, (ii) signature based object detection. 

\subsubsection{AS for MS classification  } \label{subsec:AS_for_MS_classification  } 

\begin{figure}
\hspace{-0.3cm}
\includegraphics[width=9.5cm]{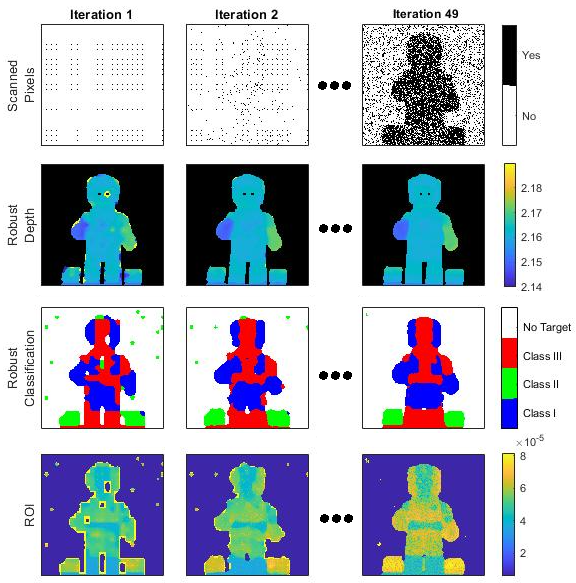}
\caption{Illustrative results of the AS algorithm with respect to iterations. The 
columns represent the process at a particular iteration and the rows represent from
top to bottom: scanned samples, robust depth estimation using spatial correlation between pixels,  robust class detection using spatial correlation between pixels and the ROI map where yellow
regions indicate important regions to sample ($\omega=1.3$, $N_s=32^2$, $t_0=1500 \mu s $, $T=1500$).} \label{fig:lego_dirty_iteration_depth_class_ROI}
\end{figure} 
The first study consider a classification task, where the sensing aims to detect pixels with target, and to classify them into $K=3$ classes based on known spectral signatures (see Fig.  \ref{fig:IRF}). Figs. \ref{fig:RMSE_lego_clean_usrsas} and  \ref{fig:RMSE_lego_dirty_usrsas} shows depth RMSEs  for negligible ($\omega=66$) and high background illumination ($\omega=1.3$), respectively. In this case, the region of interest is large and occupies most of the HR image, which reduces the benefit of the proposed targeting approach. Nonetheless,  Figs. \ref{fig:RMSE_lego_clean_usrsas} and  \ref{fig:RMSE_lego_dirty_usrsas} show a noticeable improvement by the proposed AS strategy especially for low SBR (e.g., In Fig.\ref{fig:RMSE_lego_dirty_usrsas} and for RMSE$=1.25$mm, we observe an improvement factor of $80$ and $18$ based on dwell or total times, respectively). Table \ref{tab:confMat_lego_200by200_clean2} shows the classification confusion matrix for the considered Lego scene for an average acquisition time of 
$75 \mu s$ per-pixel and per wavelength (less than 3 seconds of total acquisition time per wavelength). This table highlights good classification results with an accuracy $\geq 0.9$ even at low SBR.
Finally,  Fig. \ref{fig:lego_dirty_iteration_depth_class_ROI} illustrates the visual performance of the algorithm on the Lego data ($\omega = 1.3$) with respect to iterations. This figure indicates that most samples are located in the Lego region as promoted by the ROI map, and that both depth and classification maps improve with iterations.

\begin{table}[] 
\centering
\begin{tabular}{lllll|l}
\cline{2-6} 
\multicolumn{1}{c|}{}  & \multicolumn{4}{c|}{Predicted classes} & \multicolumn{1}{c|}{Recall} \\ 
\hline
\multicolumn{1}{|c|}{} & \multicolumn{1}{c|}{\textbf{4950}}  & \multicolumn{1}{c|}{\textbf{45}}  & \multicolumn{1}{c|}{\textbf{679}}  & \multicolumn{1}{c|}{\textbf{102}}  & \multicolumn{1}{c|}{\multirow{2}{*}{85.7$\%$} }  \\ 
\multicolumn{1}{|c|}{True classes}  & \multicolumn{1}{c|}{12.38$\%$}  & \multicolumn{1}{c|}{0.11$\%$}  & \multicolumn{1}{c|}{1.7$\%$}  & \multicolumn{1}{c|}{0.26$\%$} & \multicolumn{1}{c|}{}\\
\cline{2-6} 
\multicolumn{1}{|c|}{} & \multicolumn{1}{c|}{\textbf{22}}  & \multicolumn{1}{c|}{\textbf{1548}}  & \multicolumn{1}{c|}{\textbf{75}}  & \multicolumn{1}{c|}{\textbf{23}}      & \multicolumn{1}{c|}{\multirow{2}{*}{92.8$\%$} }  \\
 \multicolumn{1}{|c|}{for}  & \multicolumn{1}{c|}{0.06$\%$}  & \multicolumn{1}{c|}{3.87$\%$}  & \multicolumn{1}{c|}{0.19$\%$}  & \multicolumn{1}{c|}{0.06$\%$} & \multicolumn{1}{c|}{} \\
\cline{2-6} 
\multicolumn{1}{|c|}{} & \multicolumn{1}{c|}{\textbf{557}}  & \multicolumn{1}{c|}{\textbf{49}}  & \multicolumn{1}{c|}{\textbf{6286}}  & \multicolumn{1}{c|}{\textbf{45}}        & \multicolumn{1}{c|}{\multirow{2}{*}{90.6$\%$} } \\
 \multicolumn{1}{|c|}{$\omega = 66$}  & \multicolumn{1}{c|}{1.39$\%$}  & \multicolumn{1}{c|}{0.12$\%$}  & \multicolumn{1}{c|}{15.72$\%$}  & \multicolumn{1}{c|}{0.11$\%$} & \multicolumn{1}{c|}{} \\
\cline{2-6} 
\multicolumn{1}{|c|}{} & \multicolumn{1}{c|}{\textbf{332}}  & \multicolumn{1}{c|}{\textbf{352}}  & \multicolumn{1}{c|}{\textbf{264}}  & \textbf{2476}                     & \multicolumn{1}{c|}{\multirow{2}{*}{72.3$\%$} }  \\
 \multicolumn{1}{|c|}{}  & \multicolumn{1}{c|}{0.83$\%$}  & \multicolumn{1}{c|}{0.88$\%$}  & \multicolumn{1}{c|}{0.66$\%$}  & \multicolumn{1}{c|}{61.68$\%$} & \multicolumn{1}{c|}{} \\
\hline
\multicolumn{1}{|c|}{\multirow{2}{*}{Precision} } &
\multicolumn{1}{c|}{\multirow{2}{*}{84.5$\%$} } & \multicolumn{1}{c|}{\multirow{2}{*}{77.6$\%$} } & \multicolumn{1}{c|}{\multirow{2}{*}{86.1$\%$} } & \multicolumn{1}{c|}{\multirow{2}{*}{99.3$\%$} }                    & \multicolumn{1}{c|}{\multirow{2}{*}{93.9$\%$} } \\
\multicolumn{1}{|c|}{}  & \multicolumn{1}{c|}{}  & \multicolumn{1}{c|}{}  & \multicolumn{1}{c|}{}  & \multicolumn{1}{c|}{} & \multicolumn{1}{c|}{} \\
\hline
\end{tabular}

\vspace{0.4cm}

\begin{tabular}{lllll|l}
\hline
\multicolumn{1}{|c|}{} & \multicolumn{1}{c|}{\textbf{5022}}  & \multicolumn{1}{c|}{\textbf{29}}  & \multicolumn{1}{c|}{\textbf{594}}  & \multicolumn{1}{c|}{\textbf{131}}  & \multicolumn{1}{c|}{\multirow{2}{*}{87$\%$} }  \\ 
\multicolumn{1}{|c|}{True classes}  & \multicolumn{1}{c|}{12.56$\%$}  & \multicolumn{1}{c|}{0.07$\%$}  & \multicolumn{1}{c|}{1.49$\%$}  & \multicolumn{1}{c|}{0.33$\%$} & \multicolumn{1}{c|}{}\\
\cline{2-6} 
\multicolumn{1}{|c|}{} & \multicolumn{1}{c|}{\textbf{24}}  & \multicolumn{1}{c|}{\textbf{1530}}  & \multicolumn{1}{c|}{\textbf{93}}  & \multicolumn{1}{c|}{\textbf{21}}      & \multicolumn{1}{c|}{\multirow{2}{*}{91.7$\%$} }  \\
 \multicolumn{1}{|c|}{for}  & \multicolumn{1}{c|}{0.06$\%$}  & \multicolumn{1}{c|}{3.82$\%$}  & \multicolumn{1}{c|}{0.23$\%$}  & \multicolumn{1}{c|}{0.05$\%$} & \multicolumn{1}{c|}{} \\
\cline{2-6} 
\multicolumn{1}{|c|}{} & \multicolumn{1}{c|}{\textbf{616}}  & \multicolumn{1}{c|}{\textbf{48}}  & \multicolumn{1}{c|}{\textbf{6231}}  & \multicolumn{1}{c|}{\textbf{42}}        & \multicolumn{1}{c|}{\multirow{2}{*}{89.8$\%$} } \\
 \multicolumn{1}{|c|}{$\omega = 1.3$}  & \multicolumn{1}{c|}{1.54$\%$}  & \multicolumn{1}{c|}{0.12$\%$}  & \multicolumn{1}{c|}{15.58$\%$}  & \multicolumn{1}{c|}{0.11$\%$} & \multicolumn{1}{c|}{} \\
\cline{2-6} 
\multicolumn{1}{|c|}{} & \multicolumn{1}{c|}{\textbf{262}}  & \multicolumn{1}{c|}{\textbf{395}}  & \multicolumn{1}{c|}{\textbf{310}}  & \textbf{24652}                     & \multicolumn{1}{c|}{\multirow{2}{*}{96.2$\%$} }  \\
 \multicolumn{1}{|c|}{}  & \multicolumn{1}{c|}{0.66$\%$}  & \multicolumn{1}{c|}{0.99$\%$}  & \multicolumn{1}{c|}{0.78$\%$}  & \multicolumn{1}{c|}{61.63$\%$} & \multicolumn{1}{c|}{} \\
\hline
\multicolumn{1}{|c|}{\multirow{2}{*}{Precision} } &
\multicolumn{1}{c|}{\multirow{2}{*}{84.8$\%$} } & \multicolumn{1}{c|}{\multirow{2}{*}{76.4$\%$} } & \multicolumn{1}{c|}{\multirow{2}{*}{86.2$\%$} } & \multicolumn{1}{c|}{\multirow{2}{*}{99.2$\%$} }                    & \multicolumn{1}{c|}{\multirow{2}{*}{93.6$\%$} } \\
\multicolumn{1}{|c|}{}  & \multicolumn{1}{c|}{}  & \multicolumn{1}{c|}{}  & \multicolumn{1}{c|}{}  & \multicolumn{1}{c|}{} & \multicolumn{1}{c|}{} \\
\cline{1-6}  
\end{tabular}
\caption{Confusion matrix of the Lego scene ($200 \times 200$ pixels) in (top) $\omega=66$ and $3$ seconds of acquisition time per wavelength ($\approx 26$ signal photons per-pixel)  and (bottom) $\omega=1.3$ and $2.55$ seconds of acquisition time per-pixel and per wavelength ($\approx 22$ signal photons per-pixel and per wavelength)) .}
\label{tab:confMat_lego_200by200_clean2}
\end{table}

\subsubsection{AS for signature based object detection} \label{subsec:AS_for_MS_target_detection} 
Being a scene dependent approach, adaptive sampling is sensitive to the structure and the distribution of objects of interest in a particular scene. So far, the objects of interest that we tested occupy an important portion of scene (where the ROI map occupies more than $50\%$ of the HR grid).
To simulate a real world scenario, where targets are usually occupying a smaller portion of the scene, the algorithm is run using the samples Lego data, but targeting a single class corresponding to the green blocks on both sides of the Lego (see top sub-figures of Fig. \ref{fig:IRF}). In addition to their small size, the green blocks have low reflectivity, hence present high uncertainty. Thanks to the ROI map, the algorithm will be able to scan this region and reduce the uncertainty measure by scanning more around that region. This task is akin to target detection where the target is a single class of interest. 
Figs. \ref{fig:RMSE_lego_clean_one_class_usrsas} and  \ref{fig:RMSE_lego_dirty_one_class_usrsas} show the resulting depth RMSEs for $\omega=66$ and  $\omega=1.3$, respectively. The latter figure reflects an interesting and challenging real-world scenario as we are most likely to be interested in locating: i) objects that occupy a small portion of a scene, ii)  with a specific spectral signature, and iii) within an adverse environment (low SBR). In that case, AS shows a significant improvement in acquisition time when compared to static strategies, reaching an improvement factor of $10$ and $200$ at RMSE $=1$mm for high and low SBR, respectively.

\section{Conclusions} \label{sec:Conclusions}  \vspace{-0.0cm}

This paper has presented a task optimized adaptive sampling framework for multispectral single-photon LiDAR data. The iterative approach samples new points to reduce the uncertainties of the parameters of interest and hence improve their estimates. We demonstrated the framework when considering two tasks: (i) a classification task where the goal is to improve the labeling of pixels based on their spectral signatures, and (ii) signature based object detection. In both cases, pixels only containing background photons are ignored to concentrate on informative pixels with target reflections.  
A new Bayesian model was proposed to perform these tasks by providing the parameter posterior distributions, which contain parameter estimates together with a measure of their uncertainties. Several experiments on simulated and real data were performed showing the clear benefit of the proposed framework when compared to static sampling strategies. More precisely, we demonstrated faster convergence of the depth estimate especially in realistic scenarios involving a high background, and/or spatially small targets of interest. We have also studied the use of different detector array sizes, together with scanning modes that can be sequential or parallel.  
As expected, it was observed that parallel scanning allows for faster acquisition. While large arrays allows faster scanning, it was also shown that small arrays allow finer sampling and hence the targets shape. Future work will investigate a new Bayesian formulation based on photon events to avoid building histograms of counts as in the current approach. Generalizing the proposed approach by adopting a recursive Bayesian model to update the parameter prior distributions will also be the subject of future work.

 
\bibliographystyle{IEEEtran}
\bibliography{biblio_all}
\end{document}